\documentclass[11pt]{article}
\usepackage[T1]{fontenc}
\usepackage{epsfig}
\usepackage{latexsym}
\usepackage{amssymb}
\usepackage{amsthm}
\usepackage{amsmath}
\usepackage{mathrsfs}
\usepackage{hyperref}
\hypersetup{
   colorlinks,
    citecolor=blue,
    filecolor=black,
    linkcolor=red,
    urlcolor=magenta
}
\hypersetup{linktocpage}

\newcommand{\R}{\mathbb{R}} \newcommand{\N}{\mathbb{N}}

\renewcommand{\d}{\mathrm{d}}

\newcommand{\dvol}{\mathrm{dVol}}

\newcommand{\scri}{{\mathscr I}}
\newcommand{\scrh}{{\mathscr H}}

\DeclareMathOperator*{\slim}{\mathrm{s}\,--\lim ~}
\newcommand{\hgamma}{\hat\Gamma}
\newcommand{\ba}{\mathbf{a}}
\newtheorem{definition}{Definition}
\newtheorem{theorem}{Theorem}
\newtheorem{proposition}{Proposition}
\newtheorem{corollary}{Corollary}
\newtheorem{lemma}{Lemma}
\newtheorem{remark}{Remark}

\begin{document}
%\selectlanguage{english}
\mbox{} \thispagestyle{empty}

\begin{center}
\bf{\Huge Conformal scattering

\vspace{0.1in}

on the Schwarzschild metric} \\

\vspace{0.1in}

{Jean-Philippe NICOLAS}\\
\small{\it{Department of Mathematics,}} \\
\small{\it{University of Brest, 6 avenue Victor Le Gorgeu,}} \\
\small{\it{29200 Brest, France.}} \\
\small{\it{Email~: Jean-Philippe.Nicolas@univ-brest.fr}}
\end{center}

{\bf Abstract.} We show that existing decay results for scalar fields on the Schwarzschild metric are sufficient to obtain a conformal scattering theory. Then we re-interpret this as an analytic scattering theory defined in terms of wave operators, with an explicit comparison dynamics associated with the principal null geodesic congruences. The case of the Kerr metric is also discussed.



\tableofcontents

\section{Introduction}

Conformal time dependent scattering originates from the combination of the ideas of R. Penrose on spacetime conformal compactification \cite{Pe1963, Pe1964, Pe1965, PeRi}, the Lax-Phillips theory of scattering \cite{LaPhi} and F.G. Friedlander's notion of radiation fields \cite{Fri1962, Fri1964, Fri1967}. The Lax-Phillips scattering theory for the wave equation is a construction on flat spacetime. It is based on a translation representer of the solution, which is re-interpreted as an asymptotic profile of the field along outgoing radial null geodesics, analogous to Friedlander's radiation field\footnote{It is interesting to note that the integral formula, obtained by Lax and Phillips, that recovers the field in terms of its scattering data, was in fact discovered by E.T. Whittaker in 1903 \cite{Whi}. This does not seem to have been known to them or to Friedlander. The Lax-Phillips theory gave Whittaker's formula its rightful interpretation as a scattering representation of the solutions of the wave equation. There is an interesting extension of this formula to plane wave spacetimes due to R.S. Ward \cite{Wa}, developed further by L.J. Mason \cite{Ma}.}. Observing this, Friedlander formulated the first version of conformal time-dependent scattering in 1980 \cite{Fri1980}. The framework was a static spacetime with a metric approaching the flat metric fast enough at infinity (like $1/r^2$) so as to ensure that the conformal spacetime has a regular null infinity (denoted $\scri$). This allowed him to construct radiation fields as traces on $\scri$ of conformally rescaled fields. The scattering theory as such was obtained by the resolution of a Goursat (characteristic Cauchy) problem on null infinity, whose data are the radiation fields. Then he went on to recover the analytically explicit aspects of the Lax-Phillips theory, in particular the translation representation of the propagator, a feature which is tied in with the staticity of the geometry\footnote{More precisely, the existence of a translation representation of the propagator is tied in with the existence of a timelike Killing vector field that extends as the null generator of null infinity.}. His ideas were taken up by J.C. Baez, I.E. Segal and Zhou Z.F. in 1989-1990 \cite{Ba1989a, Ba1989b, Ba1990, BaSeZho1990, BaZho1989} to develop conformal scattering theories on flat spacetime for non linear equations. Note that the resolution of the characteristic Cauchy problem was the object of a short paper by L. Hörmander in 1990 \cite{Ho1990}, in which he described a method of resolution based entirely on energy estimates and weak compactness, for the wave equation on a general spatially compact spacetime.

Friedlander himself came back to conformal scattering just before his death in a paper published posthumously in 2001 \cite{Fri2001}. It is on the whole quite surprising that his idea did not entail more active research in the domain. It is even more puzzling that the research it did entail remained strictly focused on static geometries. In fact, the observation that a complete scattering theory in the physical spacetime, amounts to the resolution of a Goursat problem on the compactified spacetime, is the door open to the development of scattering theories on generic non stationary geometries. Probably Friedlander's wish to recover all the analytic richness of the Lax-Phillips theory prevented him from pushing his theory this far. However, the door being open, somebody had to go through it one day. This was done in 2004 by L.J. Mason and the author in \cite{MaNi2004}, a paper in which a conformal scattering theory was developed for scalar waves\footnote{The treatment of the wave equation was not completed in this paper, the additional ingredients required can be found in another work by the same authors, dealing with the peeling of scalar fields, published in 2009 \cite{MaNi2009}.}, Dirac and Maxwell fields, on generically non stationary asymptotically simple spacetimes. A conformal scattering theory for a non linear wave equation on non stationary backgrounds was then obtained by J. Joudioux in 2012 \cite{Jo2012}.

The purpose of the present work is to show how existing decay results can be used to obtain conformal scattering constructions on black hole backgrounds. We treat the case of the wave equation on the Schwarzschild metric, for which the analytic scattering theory is already known (see J. Dimock and B.S. Kay in 1985-1987 \cite{Di1985, DiKa1986, DiKa1987}). The staticity of the exterior of the black hole gives a positive definite conserved quantity on spacelike slices, which can be extended to the conformally rescaled spacetime~; the known decay results (we use those of M. Dafermos and I. Rodnianski, see for example their lecture notes \cite{DaRoLN}) are then enough to obtain a complete scattering theory. It is in some sense unsatisfactory to use decay results, because they require a precise understanding of the trapping by the photon sphere, which is much more information than is needed for a scattering theory. However, such results should by nature be fairly robust under small perturbations. So the conformal scattering theories on stationary black hole backgrounds obtained using them can in principle be extended to non stationary perturbations. Not that this is at all trivial. This work is to be considered a first step in the developent of conformal scattering theories on black hole backgrounds, to be followed by extensions to other equations and to more general, non stationary situations.

The paper is organized as follows. Section \ref{GeomFrame} contains the description of the geometrical framework for the case of the wave equation on the Schwarzschild metric. We describe the conformal compactification of the geometry and the corresponding rescaling of the wave equation. In section \ref{EnIdent}, we derive the main energy estimates on the compactified spacetime. Section \ref{Scattering} is devoted to the conformal scattering construction and to its re-interpretation in terms of wave operators associated to a comparison dynamics. This type of structure, contrary to the translation representation, would survive in a non stationary situation (see \cite{MaNi2004} for an analogous construction on non stationary asymptotically simple spacetimes). This re-interpretation concerns the most difficult aspects of analytic scattering theory~: the existence of inverse wave operators and asymptotic completeness. For the existence of direct wave operators, which is the easy part, we keep the analytic approach using Cook's method~; this is explained in appendix \ref{AppendixCook}. The reason for this choice is the simplicity of the method and its easy entendibility to fairly general geometries, using a geometric transport equation as comparison dynamics, provided we have a precise knowledge of the asymptotic behaviour of the metric and good uniform energy estimates (which are in any case crucial for developing a conformal scattering theory). Some technical aspects of the resolution of the Goursat problem on the conformal boundary, which is at the core of the conformal scattering theory, are explained in appendix \ref{HormGP}. Section \ref{Kerr} is devoted to remarks concerning the extension of these results to the Kerr metric and some concluding comments. Since the first version of this work, this last section has been entirely re-written in order to take the new results by M. Dafermos, I. Rodnianski and Y. Shlapentokh-Rothman \cite{DaRoShla} into account.

{\bf Notations and conventions.} Given a smooth manifold $M$ without boundary, we denote by ${\cal C}^\infty_0 (M)$ the space of smooth compactly supported scalar functions on $M$ and by ${\cal D}' (M)$ its topological dual, the space of distributions on $M$.

Concerning differential forms and Hodge duality, following R. Penrose and W. Rindler \cite{PeRi}, we adopt the following convention~: on a spacetime $({\cal M},g)$ (i.e. a $4$-dimensional Lorentzian manifold that is oriented and time-oriented), the Hodge dual of a $1$-form $\alpha$ is given by
\[ (*\alpha)_{abc} = e_{abcd} \alpha^d\, , \]
where $e_{abcd}$ is the volume form on $({\cal M} , g)$, which in this paper we simply denote $\dvol$. We shall use two important properties of the Hodge star~:
\begin{itemize}
\item given two $1$-forms $\alpha$ and $\beta$, we have
\begin{equation} \label{HStarP1}
\alpha \wedge * \beta = -\frac{1}{4} \alpha_a  \beta^a\, \dvol  \, ;
\end{equation}
\item for a $1$-form $\alpha$ that is differentiable,
\begin{equation} \label{HStarP2}
\d * \alpha =  -\frac{1}{4} (\nabla_a  \alpha^a ) \dvol  \, .
\end{equation}
\end{itemize}

\begin{remark}[Conformal and analytic scattering]
Throughout this work, we shall talk about analytic and conformal scattering as two different approaches to scattering theory. In most cases, we mean that the former is based on spectral techniques and the latter relies on a conformal compactification. The truly significant difference however is that conformal scattering understands the scattering construction as the resolution of a Goursat problem on the conformal boundary, described as a finite hypersurface, whereas analytic scattering sees the scattering channels as asymptotic regions.
\end{remark}

\section{Geometrical framework} \label{GeomFrame}

The Schwarzschild metric is given on $\R_t \times ]0,+\infty [_r \times S^2_\omega$ by
\[ g = F \d t^2 - F^{-1} \d r^2 - r^2 \d \omega^2 \, ,~ F = F(r) = 1 -\frac{2M}{r} \, , \]
where $\d \omega^2$ (also denoted $e_{S^2}$ below) is the euclidean metric on $S^2$ and $M>0$ is the mass of the black hole. We work on the exterior of the black hole $\{ r>2M \}$, which is the only region of spacetime perceived by static observers at infinity (think for instance of a distant telescope pointed at the black hole). Introducing the Regge-Wheeler coordinate $r_* = r + 2M \log (r-2M)$, such that $\d r = F \d r_*$, the metric $g$ takes the form
\[ g = F (\d t^2 - \d r_*^2 ) - r^2 \d \omega^2 \, .\]
The Schwarzschild metric has a four-dimensional space of global Killing vector fields, generated by
\begin{equation} \label{KVF}
K:=\partial_t \, ,~ X:=\sin \varphi \, \partial_\theta + \cot \theta \cos \varphi \, \partial_\varphi \, ,~Y:=\cos \varphi \, \partial_\theta - \cot \theta \sin \varphi \, \partial_\varphi \, ,~Z:=\partial_\varphi \, ,
\end{equation}
which are the timelike (outside the black hole) Killing vector field $\partial_t$ and the three generators of the rotation group. Some other essential vector fields are the principal null vector fields (the vectors we give here are ``unnormalized'', they are not the first two vectors of a normalized Newman-Penrose tetrad)
\begin{equation} \label{PND}
l = \partial_t + \partial_{r_*}  \, ,~ n = \partial_t - \partial_{r_*} \, .
\end{equation}

We perform a conformal compactification of the exterior region using the conformal factor $\Omega = 1/r$, i.e. we put
\[ \hat{g} = \Omega^2 g \, .\]
To express the rescaled Schwarzschild metric, we use coordinates $u = t-r_*$, $R=1/r$, $\omega$~:
\begin{equation} \label{ghatu}
\hat{g} = R^2 (1-2MR) \d u^2 - 2 \d u \d R - \d \omega^2 \, .
\end{equation}
The inverse metric is
\begin{equation} \label{InvRescSchwaMet}
\hat{g}^{-1} = - \partial_u \otimes \partial_R - \partial_R \otimes \partial_u - R^2 (1-2MR) \partial_R\otimes \partial_R - e^{-1}_{S^2} \, .
\end{equation}
The non-zero Christoffel symbols for $\hat{g}$ in the coordinates $u,R,\omega$ are~:
\begin{gather*}
\hgamma^0_{00} = R (1-3MR) \, ,~ \hgamma^1_{00} = R^3 (1-2MR)(1-3MR) \, ,~ \hgamma^1_{01} = -R(1-3MR) \, , \\
\hgamma^2_{33} = -\sin \theta \cos \theta \, ,~ \hgamma^3_{23} = \cot \theta \, .
\end{gather*}
If we use the coordinates $(t,r,\theta , \varphi )$, we get instead (still for the metric $\hat{g}$)
\begin{gather*}
\hgamma^0_{01} = \frac{3M-r}{r(r-2M)} \, ,~ \hgamma^1_{00} = \frac{(r-2M)(3M-r)}{r^3} \, ,~ \hgamma^1_{11} = \frac{M-r}{r(r-2M)} \, , \\
\hgamma^2_{33} = -\sin \theta \cos \theta \, ,~ \hgamma^3_{23} = \cot \theta \, ,
\end{gather*}
the others being zero.

\begin{figure}[ht] %  figure placement: here, top, bottom, or page
\centering
\includegraphics[width=4in]{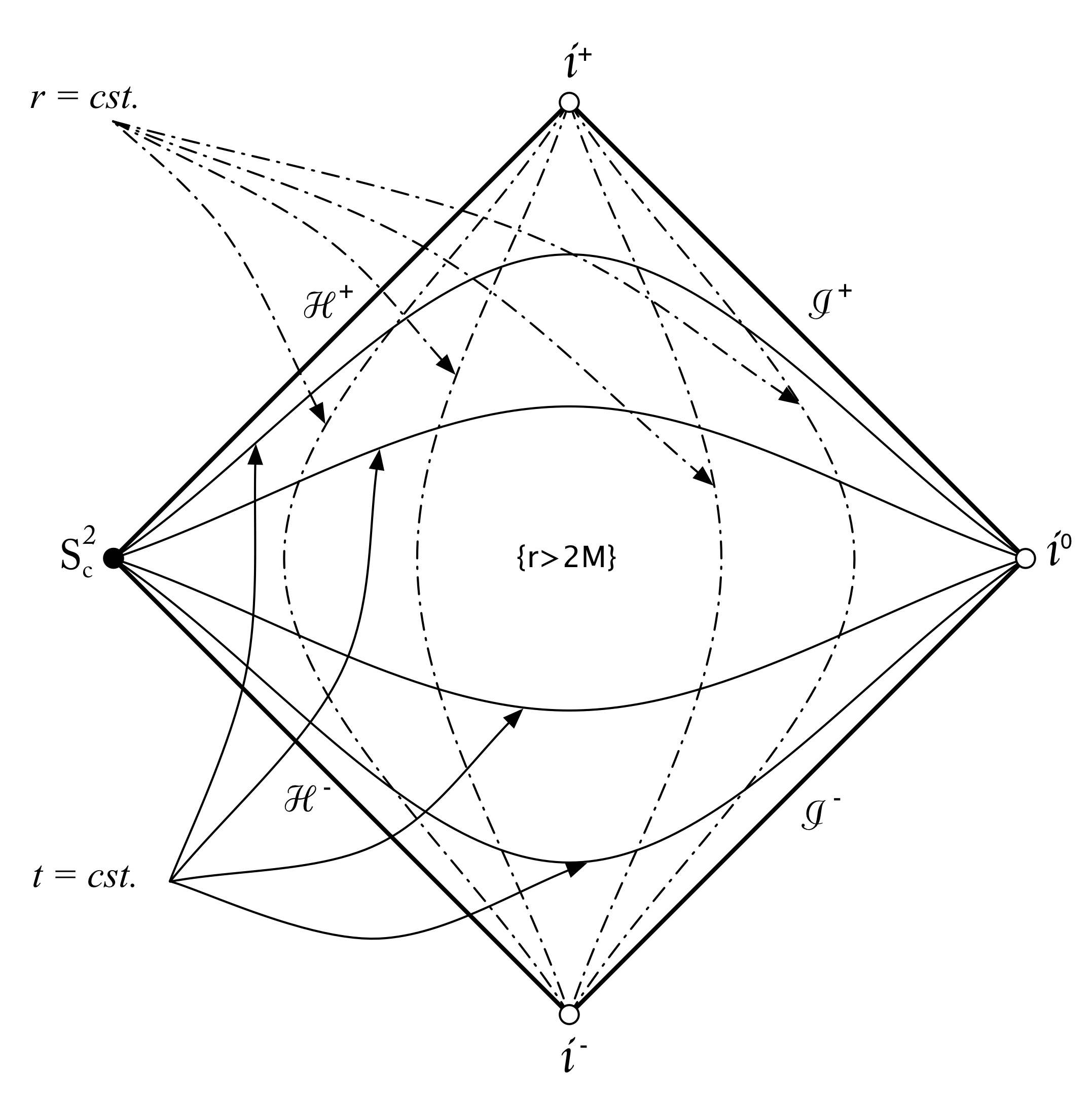}
\caption{Carter-Penrose diagram of the conformal compactification of the exterior of the black hole.} \label{PenD}
\end{figure}

Future null infinity $\scri^+$ and the past horizon $\scrh^-$ are null hupersurfaces of the rescaled spacetime
\[ \scri^+ = \R_u \times \{ 0\}_R \times S^2_\omega \, ,~ \scrh^- = \R_u \times \{ 1/2M \}_R \times S^2_\omega \, .\]
If instead of $u,R,\omega$ we use the coordinates $v=t+r_*,R,\omega$, the metric $\hat{g}$ takes the form
\begin{equation} \label{ghatv}
\hat{g} = R^2 (1-2MR) \d v^2 + 2 \d v \d R - \d \omega^2 \, .
\end{equation}
In these coordinates we have access to past null infinity $\scri^-$ and the future horizon $\scrh^+$ described as the null hypersurfaces
\[ \scri^- = \R_v \times \{ 0\}_R \times S^2_\omega \, ,~ \scrh^+ = \R_v \times \{ 1/2M \}_R \times S^2_\omega \, .\]
The compactification is not complete~; spacelike infinity $i^0$ and the timelike infinities $i^\pm$ remain at infinity for $\hat{g}$. The crossing sphere $S^2_\mathrm{c}$, which is the boundary of all level hypersurfaces of $t$ outside the black hole and the place where the future and past horizons meet, is not at infinity but it is not described by the coordinate systems $\{u,R,\omega \}$ and $\{v,R,\omega \}$~; it is the only place in $\{ r\geq 2M \} \cup \scri^\pm$ where $\partial_t$ vanishes. See Figure \ref{PenD} for a Carter-Penrose diagram of the compactified exterior.

A crucial feature of the conformal compactification using the conformal factor $1/r$ is that it preserves the symmetries~: the vector fields \eqref{KVF} are still Killing for $\hat{g}$. In particular, the vector field $\partial_t$ becomes $\partial_u$ in the $(u,R,\omega )$ coordinate system, respectively $\partial_v$ in the $(v,R,\omega )$ coordinate system~; thus it extends as the future-oriented null generator of null infinities $\scri^\pm$ and the future and past horizons $\scrh^\pm$.

\begin{figure}[ht] %  figure placement: here, top, bottom, or page
\centering
\includegraphics[width=4in]{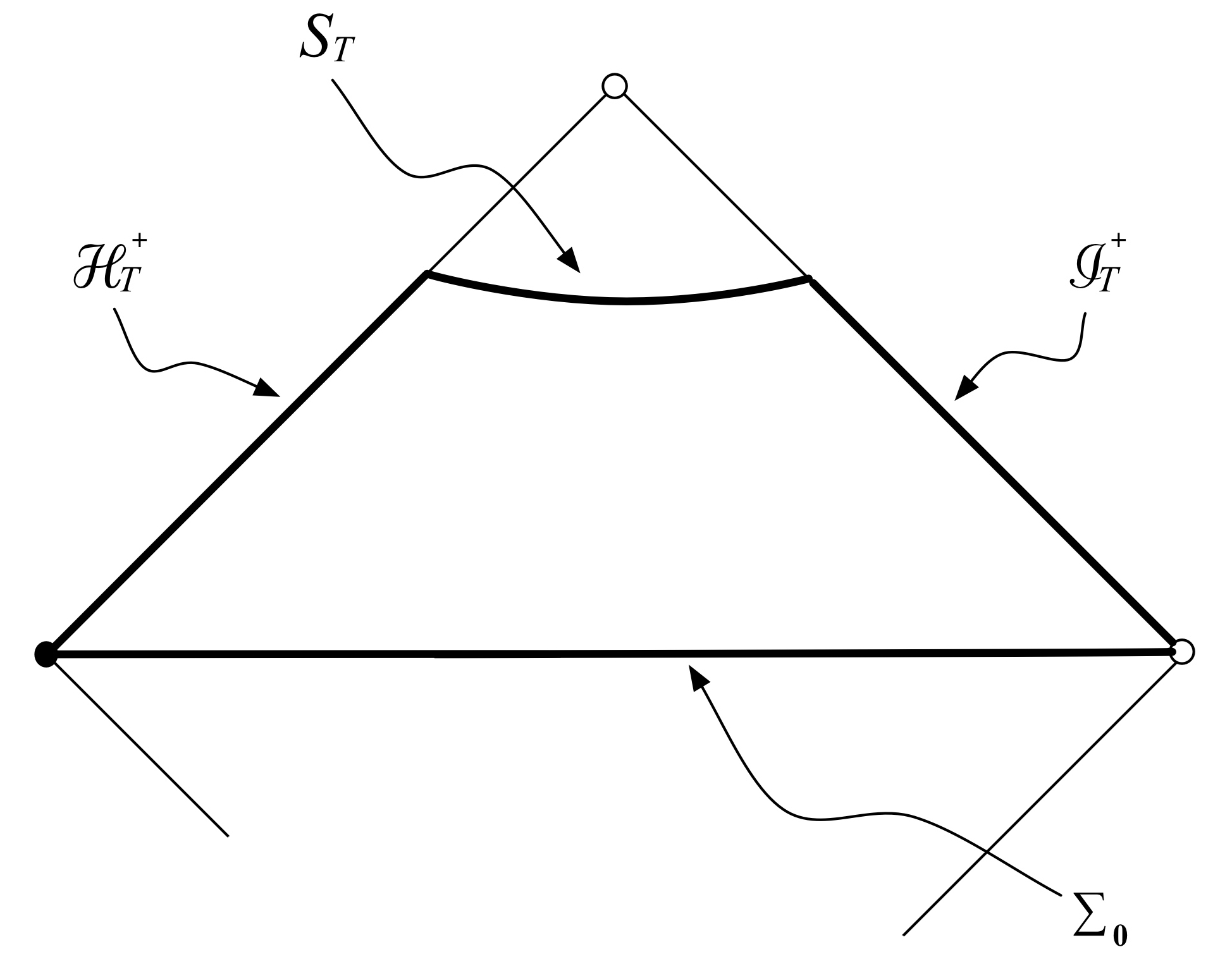}
\caption{The main hypersurfaces represented on the compactified exterior.}  \label{3surface}
\end{figure}

We shall denote by $\cal M$ the exterior of the black hole, ${\cal M}=\R_t \times ]2M , +\infty [_r \times S^2$, and by $\bar{\cal M}$ its conformal compactification, i.e.
\[ \bar{\cal M} = {\cal M} \cup \scri^+ \cup \scrh^+ \cup \scri^- \cup \scrh^- \cup S^2_c \, . \]
\begin{remark}
The constructions of the horizons and of null infinities are of a very different nature. Understanding the horizons as smooth null hypersurfaces of the analytically extended Schwarzschild exterior only requires a change of coordinates, for instance the advanced and retarded Eddington-Finkelstein coordinates $(u,R,\omega)$ and $(v,R,\omega)$. For the construction of null infinities however, the conformal rescaling is necessary and $\scri^\pm$ are boundaries of the exterior of the black hole endowed with the metric $\hat{g}$, not of the physical exterior $({\cal M},g)$.
\end{remark}
The main hypersurfaces that we shall use in this paper are the following~:
\begin{eqnarray}
\Sigma_t &=& \{ t \} \times \Sigma \, ,~ \Sigma = ]2M , +\infty [_r \times S^2_\omega = \R_{ r_*} \times S^2_\omega \, , \label{Sigt} \\
S_T &=& \left\{ (t,r_* , \omega) \in \R \times \R \times S^2 \, ;~ t = T+ \sqrt{1+r_*^2} \right\} \label{ST} \, , \\
\scri^+_T &=& \scri^+ \cap \{ u \leq T\} = ]-\infty , T]_u \times \{ 0 \}_R \times S^2_\omega \, , \label{scriT} \\
\scrh^+_T &=& S^2_{\mathrm{c}} \cup (\scrh^+ \cap \{ v \leq T\} ) = S^2_{\mathrm{c}} \cup (]-\infty , T]_v \times \{ 1/2M \}_R \times S^2_\omega ) \, . \label{scrhT}
\end{eqnarray}
For $T>0$, the hypersurfaces $\Sigma_0$, $\scrh^+_T$, $S_T$ and $\scri^+_T$ form a closed --- except for the part where $\scri^+$ and $\Sigma_0$ touch $i^0$ --- hypersurface on the compactified exterior (see Figure \ref{3surface}). We make such an explicit choice for the hypersurface $S_T$ for the sake of clarity but it is not strictly necessary, all that is required of $S_T$ is that it is uniformly spacelike for the rescaled metric, or even achronal, and forms a closed hypersurface with $\Sigma_0$, $\scrh^+_T$, and $\scri^+_T$. 

The scalar curvature of the rescaled metric $\hat{g}$ is
\[ \mathrm{Scal}_{\hat{g}} = 12MR \, .\]
So $\phi \in {\cal D}' (\R_t \times ]0,+\infty [_r \times S^2_\omega)$ satisfies
\begin{equation} \label{WEqPhys}
\square_g \phi =0 
\end{equation}
if and only if $\hat{\phi} = \Omega^{-1} \phi$ satisfies
\begin{equation} \label{WEqResc}
(\square_{\hat{g}}  + 2MR ) \hat{\phi} =0 \, .
\end{equation}
By the classic theory of hyperbolic partial differential equations (see Leray \cite{Le1953}), for smooth and compactly supported initial data $\hat{\phi}_0$ and $\hat{\phi}_1$ on $\Sigma_0$, we have the following properties~:
\begin{itemize}
\item there exists a unique $\hat{\phi} \in {\cal C}^\infty ({\cal M})$ solution of \eqref{WEqResc} such that
\[ \hat{\phi} \vert_{\Sigma_0} = \hat{\phi}_0 \mbox{ and } \partial_t \hat{\phi} \vert_{\Sigma_0} = \hat{\phi}_1 \, ,\]
\item $\hat{\phi}$ extends as a smooth function on $\bar{\cal M}$ and therefore has a smooth trace on $\scrh^\pm \cup \scri^\pm$.
\end{itemize}
The D'Alembertians for the metrics $g$ and $\hat{g}$ have the following expressions in variables $(t, r_*, \omega)$~:
\begin{eqnarray*}
\square_g &=& \frac{1}{F} \left( \frac{\partial^2}{\partial t^2} - \frac{1}{r^2} \frac{\partial}{\partial r_*} r^2 \frac{\partial}{\partial r_*} \right) - \frac{1}{r^2} \Delta_{S^2} \, ,\\
\square_{\hat{g}} &=& \frac{r^2}{F} \left( \frac{\partial^2}{\partial t^2} - \frac{\partial^2}{\partial r_*^2} \right)
 - \Delta_{S^2} \, .
\end{eqnarray*}
The volume forms associated with $g$ and $\hat{g}$ are
\begin{eqnarray}
\dvol_g &=& r^2 \sin \theta \d t \wedge \d r \wedge \d \theta \wedge \d \varphi = r^2 \d t \wedge \d r \wedge \d^2 \omega = r^2 F \d t \wedge \d r_* \wedge \d^2 \omega \, ,\\
\dvol_{\hat{g}} &=& \Omega^4 \dvol_g = R^2\d t \wedge \d r \wedge \d^2 \omega = R^2 F \d t \wedge \d r_* \wedge \d^2 \omega \, ,
\end{eqnarray}
$\d^2 \omega$ being the euclidean area element on $S^2$.

\section{Energy identities} \label{EnIdent}

The usual stress-energy tensor for the wave equation is not conformally invariant. We have therefore two possible approaches to establish energy identities or inequalities.
\begin{enumerate}
\item Work with the rescaled quantities $\hat{\phi}$ and $\hat{g}$. The main advantage is that for all $T>0$, the four hypersurfaces $\Sigma_0$, $\scrh^+_T$, $S_T$ and $\scri^+_T$ are finite hypersurfaces in our rescaled spacetime (except for the part of $\Sigma_0$ and $\scri^+$ near $i^0$, but we will work with solutions supported away from $i^0$ to establish our energy identities). However, we encounter a problem of a different kind~: equation \eqref{WEqResc} does not admit a conserved stress-energy tensor. Fortunately, it turns out that if we use the stress-energy tensor for the wave equation on the rescaled spacetime, and contract it with $\partial_t$, the error term is a divergence. Therefore, we recover an exact conservation law.
\item Work with the physical quantities $\phi$ and $g$. We have an immediate conserved stress energy tensor associated with the equation. The drawback here is that $\scri$ is at infinity. So we must use our conservation law to get energy identities on finite closed hypersurfaces, then take the limit of these identities as some parts of the hypersurfaces approach $\scri$.
\end{enumerate}
Both methods are in principle absolutely fine. We choose the first one since, thanks to the stationarity of Schwarzschild's spacetime, it gives energy identities in a more direct manner\footnote{We will still however work with both the rescaled and the physical field when comparing our energy norms with those used by other authors. Of course the indices of vectors and $1$-forms will have to be raised and lowered using the rescaled metric $\hat{g}$ when working with rescaled quantities and using the physical metric $g$ when working with unrescaled quantities.}.

By the finite propagation speed, we know that for smooth compactly supported data on $\Sigma_0$, i.e. supported away from $i^0$, the associated solution of \eqref{WEqResc} vanishes in a neighbourhood of $i^0$. For such solutions, the singularity of the conformal metric at $i^0$ can be ignored and we obtain energy identities for all $T>0$ between the hypersurfaces  $\Sigma_0$, $\scrh^+_T$, $S_T$ and $\scri^+_T$. Then we show, using known decay results, that the energy flux through $S_T$ tends to zero as $T\rightarrow +\infty$. This yields an energy identity between $\Sigma_0$, $\scrh^+$ and $\scri^+$, which carries over by density to initial data in a Hilbert space on $\Sigma_0$ (see section \ref{EnEstTInfinite} for details).

\subsection{Conserved energy current for the rescaled field}

The stress-energy tensor for the wave equation associated with $\hat{g}$ is given by
\begin{equation} \label{SET}
\hat{T}_{ab}= \hat\nabla_a \hat\phi \hat\nabla_b \hat\phi - \frac12 \langle \hat\nabla \hat\phi \, ,~ \hat\nabla \hat\phi \rangle_{\hat{g}} \, \hat{g}_{ab}\, .
\end{equation}
When $\hat\phi$ is a solution of \eqref{WEqResc}, the divergence of $\hat{T}$ is
\[ \hat\nabla^a \hat{T}_{ab} = (\square_{\hat{g}} \hat\phi ) \hat\nabla_b \hat\phi = -2MR \hat\phi \hat\nabla_b \hat\phi \, .\]
The energy current $1$-form associated with static observers is obtained by contracting $\hat{T}$ with the timelike Killing vector $K=\partial_t$~:
\[ \hat{J}_a = K^b \hat{T}_{ab} \, .\]
This is not conserved since
\begin{equation} \label{DivCurrent}
\hat\nabla^a \hat{J}_a = -2MR \hat\phi \partial_t \hat\phi \, .
\end{equation}
Putting
\[ V = MR\hat\phi^2 \partial_t \, ,\]
it is easy to see that
\[ 2MR \hat\phi \partial_t \hat\phi = \mathrm{div} V \, .\]
Indeed
\[ \mathrm{div} V = \hat\nabla_a V^a = \frac{\partial}{\partial t} \left( MR \hat\phi^2 \right) + \hgamma^\ba_{\ba 0} V^0 \]
and in the coordinate system $(t,r,\theta,\varphi)$, all the Christoffel symbols $\hgamma^\ba_{\ba 0}$ are zero. So \eqref{DivCurrent} can be written as an exact conservation law
\begin{equation} \label{ConsLaw}
\hat\nabla_a \left( \hat{J}^a + V^a \right) =0 \, ,~ \mbox{with } V = MR\hat\phi^2 \partial_t \, .
\end{equation}
\begin{remark}
The vector $V$ is causal and future oriented on $\bar{\cal M}$, timelike on $\cal M$, and the stress-energy tensor $\hat{T}_{ab}$ satisfies the dominant energy condition. Therefore, the energy flux across achronal hypersurfaces will be non negative and that across spacelike hypersurfaces will be positive definite. We will observe these properties on the explicit expressions of the fluxes that we calculate in the next section.
\end{remark}

\subsection{Energy identity up to $S_T$} \label{EnIdST}

The conservation law \eqref{ConsLaw} gives an exact energy identity between the hypersurfaces $\Sigma_0$, $\scrh^+_T$, $S_T$ and $\scri^+_T$, for solutions of the rescaled equation associated with smooth and compactly supported initial data. We denote by $\hat{\cal E}_{\partial_t , S}$ the rescaled energy flux, associated with $\partial_t$, across an oriented hypersurface $S$, i.e.\footnote{The factor $-4$ comes form the identity \eqref{HStarP2} applied to $\hat{J}_a + V_a$, i.e.
\[\d * ((\hat{J}_a + V_a )\d x^a) = -(1/4) \nabla_a (\hat{J}^a + V^a ) \dvol \, .\]}
\begin{equation} \label{RescEnS}
\hat{\cal E}_{\partial_t, S} = -4 \int_{S} * (\hat{J}_a + V_a )\d x^a \, .
\end{equation}
For any $T>0$, we have
\begin{equation} \label{EnIdentityT}
\hat{\cal E}_{\partial_t, \Sigma_0} = \hat{\cal E}_{\partial_t, \scri^+_T} + \hat{\cal E}_{\partial_t, \scrh^+_T} + \hat{\cal E}_{\partial_t, S_T} \, .
\end{equation}
\begin{remark}
The property \eqref{HStarP1} of the Hodge star gives us an easy way to express the energy flux across an oriented $3$-surface $S$
\[ \hat{\cal E}_{\partial_t, S} = -4\int_{S} * (\hat{J}_a + V_a )\d x^a = \int_S (\hat{J}_a+V_a)\hat{N}^a \, \hat{L}\lrcorner \dvol_{\hat{g}} \, ,\]
where $\hat{L}$ is a vector field transverse to $S$ and compatible with the orientation of the hypersurface, and $\hat{N}$ is the normal vector field to $S$ such that $\hat{g} (\hat{L},\hat{N})=1$.
\end{remark}

On $\Sigma_0$, we take
\[ \hat{L}= \frac{r^2}{F} \partial_t \, ,~\hat{N} = \partial_t \, .\]
On $\scri^+$, we take for $\hat{L}$ the future-oriented null vector $\hat{L}_{\scri^+} =-\partial_R$ in coordinates $u,R,\omega$. The vector field $-\partial_R$ in the exterior of the black hole is equal to $r^2 F^{-1} l$, with $l$ being the first principal null vector field given in \eqref{PND}, and extends smoothly to $\scri^+$~:
\[ \hat{L}_{\scri^+} = \left. r^2 F^{-1} l \right\vert_{\scri^+} \, .\]
On $\scrh^+$, we choose $\hat{L}_{\scrh^+} = \partial_R$ (in coordinates $v,R,\omega$), i.e.
\[ \hat{L}_{\scrh^+} = \left. r^2 F^{-1} n \right\vert_{\scrh^+}\, ,\]
where $n$ is the second principal null vector field in \eqref{PND}. On both $\scri^+$ and $\scrh^+$, we therefore have $\hat{N} = \partial_t$ (i.e. $\partial_v$ on $\scrh^+$ and $\partial_u$ on $\scri^+$). Since $V \propto \partial_t$ and on $\scri$ and $\scrh$ the vector field $\partial_t$ is null, we have $\hat{g} (V,\hat{N})=0$. The energy identity \eqref{EnIdentityT} reads
\begin{gather}
\int_{S_T} ((\hat{J}_a+V_a)\hat{N}^a) \, \hat{L}\lrcorner \dvol_{\hat{g}} + \int_{\scri^+_T} (\hat{J}_a K^a) \, \hat{L}_{\scri^+} \lrcorner \dvol_{\hat{g}} + \int_{\scrh^+_T} (\hat{J}_a K^a) \, \hat{L}_{\scrh^+} \lrcorner \dvol_{\hat{g}} \nonumber \\
= \int_{\Sigma_0} ((\hat{J}_a+V_a)K^a) \, r^2 F^{-1} \partial_t \lrcorner \dvol_{\hat{g}} \, . \label{EIT}
\end{gather}
We calculate the explicit expressions of the energy fluxes through $\scri^+_T$, $\scrh^+_T$ and $\Sigma_0$~:
\begin{eqnarray*}
\hat{\cal E}_{\partial_t, \Sigma_0} &=& \int_{\Sigma_0} (\hat{J}_a+V_a)K^a \, r^2 F^{-1} \partial_t \lrcorner \dvol_{\hat{g}} \\
&=& \frac12 \int_{\Sigma_0} \left( (\partial_t \hat\phi )^2 + (\partial_{r_*} \hat\phi )^2 + R^2 F \vert \nabla_{S^2} \hat\phi \vert^2 + 2 MFR^3\hat\phi^2 \right) \d r_* \d^2 \omega \, ; \\
\hat{\cal E}_{\partial_t, \scri^+_T} &=& \int_{\scri^+_T} \hat{J}_a K^a \hat{L}_{\scri^+} \lrcorner \dvol_{\hat{g}} = \int_{\scri^+_T} (\hat\nabla_K \hat\phi )^2 \hat{L}_{\scri^+} \lrcorner \dvol_{\hat{g}} \\
&=& \int_{\scri^+_T} (\partial_u (\hat\phi \vert_{\scri^+} ) )^2 \d u \d^2 \omega\, ; \\
\hat{\cal E}_{\partial_t, \scrh^+_T} &=& \int_{\scrh^+_T} \hat{J}_a K^a \hat{L}_{\scrh^+} \lrcorner \dvol_{\hat{g}} = \int_{\scrh^+_T} (\hat\nabla_K \hat\phi )^2 \hat{L}_{\scrh^+} \lrcorner \dvol_{\hat{g}} \\
&=& \int_{\scrh^+_T} (\partial_v (\hat\phi \vert_{\scrh^+}) )^2 \d v \d^2 \omega \, .
\end{eqnarray*}
We observe that the first flux defines a positive definite quadratic form and the two others non-negative quadratic forms.

We now calculate the flux through $S_T$. To this purpose, we make explicit choices of vectors $\hat{L}$ and $\hat{N}$ on $S_T$. Let us denote
\[ \Psi (t,r_*,\omega ) = t - \sqrt{1+r_*^2} \, ,\]
so the hypersurface $S_T$ is
\[ S_T = \{ (t,r_*,\omega) \, ; ~\Psi (t,r_*,\omega ) = T \} \, .\]
A co-normal to $S_T$ is given by
\[ N_a \d x^a = \d \Psi = \d t - \frac{r_*}{\sqrt{1+r_*^2}} \d r_* \]
and the associated normal vector field is
\[ \hat{N}^a = \hat{g}^{ab} N_b \, ,~ \mbox{i.e. } \hat{N}^a \frac{\partial}{\partial x^a} = r^2 F^{-1} \left( \frac{\partial}{\partial t} + \frac{r_*}{\sqrt{1+r_*^2}} \frac{\partial}{\partial r_*} \right) \, .\]
For the transverse vector $\hat{L}$, we can take
\[ \hat{L}^a \frac{\partial}{\partial x^a} = \frac{1+r_*^2}{1+2r_*^2} \left( \frac{\partial}{\partial t} - \frac{r_*}{\sqrt{1+r_*^2}} \frac{\partial}{\partial r_*} \right) \, ,\]
which is future-oriented and satisfies $\hat{L}_a \hat{N}^a =1$. We can now calculate the energy flux through $S_T$. First we have
\[ (\hat{J}_a+V_a )\hat{N}^a = MR \hat\phi^2 + \frac{r^2}{2F} \left( (\partial_t \hat\phi )^2 + (\partial_{r_*} \hat\phi )^2 + \frac{2r_*}{\sqrt{1+r_*^2}} \partial_t \hat\phi \partial_{r_*} \hat\phi + R^2F \vert \nabla_{S^2} \hat\phi \vert^2 \right) \, .\]
The contraction of $\hat{L}$ into the volume form for $\hat{g}$ is as follows
\[ \hat{L}\lrcorner \dvol_{\hat{g}} = \frac{1+r_*^2}{1+2r_*^2} R^2 F \sin \theta \left( \d r_* \wedge \d \theta \wedge \d \varphi + \frac{r_*}{\sqrt{1+r_*^2}} \d t \wedge \d \theta \wedge \d \varphi \right) \, .\]
On $S_T$, we have
\[ \d t = \frac{r_*}{\sqrt{1+r_*^2}} \d r_* \, ,\]
and therefore
\begin{eqnarray*}
\hat{L}\lrcorner \dvol_{\hat{g}} \vert_{S_T} &=& \frac{1+r_*^2}{1+2r_*^2} R^2 F \sin \theta \left( 1+ \frac{r_*^{2}}{1+r_*^2} \right) \d r_* \wedge \d \theta \wedge \d \varphi \\
&=& R^2 F \sin \theta \d r_* \wedge \d \theta \wedge \d \varphi \, .
\end{eqnarray*}
So we obtain
\begin{eqnarray}
\hat{\cal E}_{\partial_t, S_T} &:=& \int_{S_T} ((\hat{J}_a+V_a)N^a) \, \hat{L}\lrcorner \dvol_{\hat{g}} \nonumber \\
&=& \int_{S_T} \bigg[ MR \hat\phi^2 + \frac{r^2}{2F} \bigg( (\partial_t \hat\phi )^2 + (\partial_{r_*} \hat\phi )^2 \nonumber \\
&& \hspace{0.3in}+ \frac{2r_*}{\sqrt{1+r_*^2}} \partial_t \hat\phi \partial_{r_*} \hat\phi + R^2F \vert \nabla_{S^2} \hat\phi \vert^2 \bigg)\bigg] R^2 F \d r_* \d^2 \omega \, . \label{FluxST}
\end{eqnarray}
This is positive definite since $\vert r_* \vert < \sqrt{1+r_*^2}$ (and degenerates asymptotically as $\vert r_* \vert \rightarrow +\infty$).

The energy fluxes across $\scri^+_T$ and $\scrh^+_T$ are increasing non negative functions of $T$ and their sum is bounded by ${\cal E}_{\Sigma_0}$, by \eqref{ConsLaw} and the positivity of ${\cal E}_{S_T}$. Therefore they admit limits as $T \rightarrow +\infty$ and these limits are $\hat{\cal E}_{\partial_t, \scri^+}$ and $\hat{\cal E}_{\partial_t, \scrh^+}$. We have the following result~:
\begin{proposition}
For smooth and compactly supported initial data on $\Sigma_0$, the energy fluxes of the rescaled solution across $\scri^+$ and $\scrh^+$ are finite and satisfy
\[ \hat{\cal E}_{\partial_t, \scri^+} + \hat{\cal E}_{\partial_t, \scrh^+} \leq \hat{\cal E}_{\partial_t, \Sigma_0} \, .\]
We have equality in the estimate above if any only if
\begin{equation} \label{LidVanishing}
\lim_{T \rightarrow +\infty} \hat{\cal E}_{\partial_t, S_T} =0 \, .
\end{equation}
\end{proposition}
\begin{remark}
In order to construct a conformal scattering theory, we merely need to prove \eqref{LidVanishing} for a dense class of data, say smooth and compactly supported. The final identity will extend to minimum regularity initial data by density. Moreover, we can allow any loss of derivatives in the proof of \eqref{LidVanishing} for smooth compactly supported data, since we do not need to prove that $\hat{\cal E}_{\partial_t, S_T}$ tends to zero uniformly in terms of the data. This is the object of subsection \ref{EnEstTInfinite}.
\end{remark}

\subsection{Function space of initial data}

The scattering theory we are about to construct will be valid for a function space of initial data defined by the finiteness of the rescaled energy $\hat{\cal E}_{\partial_t, \Sigma_0}$. The analytic scattering theory constructed in \cite{Di1985} was valid for a function space of initial data defined by the finiteness of the energy of the physical field. It is interesting to notice that although the stress-energy tensor is not conformally invariant, the physical energy and the rescaled energy on $\Sigma_0$ are the same. Therefore, the function space of initial data for our conformal scattering theory is the same as in the analytic scattering theory of Dimock. Let us prove this.

Consider the stress-energy tensor for the wave equation on the Schwarzschild metric
\begin{equation} \label{PhysicalSET}
T_{ab} = \nabla_a \phi \nabla_b \phi - \frac12 \langle \nabla \phi \, ,~ \nabla \phi \rangle_g g_{ab} \, ,
\end{equation}
which satisfies
\[ \nabla^a T_{ab} =0 \]
for $\phi$ solution to the wave equation. The physical energy current $1$-form associated with static observers is
\[ J_a = K^b T_{ab} \]
where $K$ is the timelike Killing vector field $K=\partial_t$. This is conserved
\[ \nabla^a J_a = 0 \, . \]
The associated energy flux through an oriented hypersurface $S$ is given by\footnote{Of course the Hodge star in equation \eqref{PhysEnS} is associated with the physical metric, whereas in the expression \eqref{RescEnS} of the rescaled energy, it is associated with the rescaled metric.}
\begin{equation} \label{PhysEnS}
{\cal E}_{\partial_t , S} = -4 \int_S * J_a \d x^a \, .
\end{equation}
Similarly to what we saw for the rescaled energy fluxes, \eqref{PhysEnS} can be expressed more explicitely as
\[ {\cal E}_{\partial_t, S} =  \int_S J_a N^a \, L\lrcorner \dvol_{g} \, ,\]
where $L$ is a vector field transverse to $S$ and compatible with the orientation of the hypersurface, and $N$ is the normal vector field to $S$ such that $g (L,N)=1$.
\begin{lemma}
The energy fluxes $\hat{\cal E}_{\partial_t, \Sigma_0}$ and ${\cal E}_{\partial_t, \Sigma_0}$ are the same.
\end{lemma}
{\bf Proof.} A direct calculation shows that the physical energy flux across $\Sigma_0$ can be expressed in terms of $\hat\phi$ as follows
\begin{eqnarray*}
{\cal E}_{\partial_t , \Sigma_0} &=& \frac12 \int_{\Sigma_0} \left( (\partial_t \hat\phi )^2 + (\partial_{r_*} \hat\phi )^2 + \frac{F}{r^2} \vert \nabla_{S^2} \hat\phi \vert^2 + \frac{FF'}{r} \hat\phi^2 \right) \d r_* \wedge \d^2 \omega \, , \\
&=& \frac12 \int_{\Sigma_0} \left( (\partial_t \hat\phi )^2 + (\partial_{r_*} \hat\phi )^2 + \frac{F}{r^2} \vert \nabla_{S^2} \hat\phi \vert^2 + F\frac{2M}{r^3} \hat\phi^2 \right) \d r_* \wedge \d^2 \omega \, ,
\end{eqnarray*}
which is exactly the expression of the rescaled energy flux $\hat{\cal E}_{\partial_t , \Sigma_0}$. \qed
\begin{definition}[Finite energy space]
We denote by $\cal H$ the completion of ${\cal C}^\infty_0 (\Sigma ) \times {\cal C}^\infty_0 (\Sigma )$ in the norm
\[ \Vert (\hat{\phi}_0 \, ,~ \hat{\phi}_1 ) \Vert_{\cal H} = \frac{1}{\sqrt{2}} \left( \int_{\Sigma} \left( (\hat\phi_1 )^2 + (\partial_{r_*} \hat\phi_0 )^2 + \frac{F}{r^2} \vert \nabla_{S^2} \hat\phi_0 \vert^2 + F\frac{2M}{r^3} \hat\phi_0^2 \right) \d r_* \wedge \d^2 \omega \right)^{1/2} \, . \]
\end{definition}
The following result is classic. Its second part can be proved by Leray's theorem combined with energy identities. Its first part may be established by either the same method or by a spectral approach (showing that the Hamiltonian for \eqref{WEqPhys} is self-adjoint on $\cal H$ as this was done in \cite{Di1985} and \cite{Ni1995}).
\begin{proposition} \label{CauchyPb}
The Cauchy problem for \eqref{WEqPhys} on $\cal M$ (and therefore also for \eqref{WEqResc}) is well-posed in $\cal H$, i.e. for any $(\hat{\phi}_0  \, ,~ \hat{\phi}_1 ) \in {\cal H}$, there exists a unique $\phi \in {\cal D}' ({\cal M})$ solution of \eqref{WEqPhys} such that~:
\[ (r \phi \, ,~ r\partial_t \phi ) \in {\cal C} (\R_t \, ;~ {\cal H}) \, ; ~ r \phi \vert_{t=0} = \hat{\phi}_0 \, ;~ r \partial_t \phi \vert_{t=0} = \hat{\phi}_1 \, . \]
Moreover, $\hat{\phi}=r\phi$ belongs to $H^1_{\mathrm{loc}} (\bar{\cal M} )$ (see Remark \ref{Hsloc1} and Definition \ref{Hsloc2} below).
\end{proposition}
\begin{remark} \label{Hsloc1}
The notation $H^s_\mathrm{loc} (\bar{\cal M})$ is a perhaps not ideal, Sobolev spaces being defined on open sets. What we mean by this notation is merely that the conformal boundary is seen as a finite boundary~: only the neighbourhoods of $i^\pm$ and $i^0$ are considered as asymptotic regions in $\cal M$. With this in mind the definition of $H^s_\mathrm{loc} (\bar{\cal M})$, $s\in [0,+\infty [$ is unambiguous and goes as follows.
\end{remark}
\begin{definition} \label{Hsloc2}
Let $s \in [0,+\infty [$, a scalar function $u$ on ${\cal M}$ is said to belong to $H^s_{\mathrm{loc}} (\bar{\cal M})$ if for any local chart $(\Omega , \zeta )$, such that $\Omega \subset \cal M$ is an open set with smooth compact boundary in $\bar{\cal M}$ (note that this excludes neighbourhoods of either $i^\pm$ or $i^0$ but allows open sets whose boundary contains parts of the conformal boundary) and $\zeta$ is a smooth diffeomorphism from $\Omega$ onto a bounded open set $U \subset \R^4$ with smooth compact boundary, we have $u \circ \zeta^{-1} \in H^s ( U )$.
\end{definition}

\subsection{Energy identity up to $i^+$ and trace operator} \label{EnEstTInfinite}

Here, we prove \eqref{LidVanishing} for smooth and compactly supported data, using the estimates obtained in M. Dafermos and I. Rodnianski \cite{DaRoLN}. Theorem 4.1 in \cite{DaRoLN} contains sufficient information~: an estimate giving decay of energy with a loss of 3 angular derivatives and one order of fall-off, as well as uniform decay estimates for more regular solutions with sufficiently fast fall-off at infinity. These are expressed in terms of quantities on the physical spacetime, i.e. unrescaled quantities. We need to make sure that they give the correct information for our energy on $S_T$, which is entirely expressed in terms of rescaled quantities~; this is not completely direct since the usual stress-energy tensor for the wave equation is not conformally invariant. We start by translating their estimates using the notations we have adopted here.

Theorem 4.1 in \cite{DaRoLN} is expressed for a spacelike hypersurface for the metric $\hat{g}$ that crosses $\scrh^+$ and $\scri^+$, i.e. an asymptotically hyperbolic hypersurface for $g$, defined by translation along $\partial_t$ of a reference asymptotically hyperbolic hypersurface. Our hypersurface $S_T$ fits in this framework. The content of the theorem is the following.
\begin{description}
\item[(i)] Consider the stress-energy tensor for the wave equation on the Schwarzschild metric~: $T_{ab}$ given by \eqref{PhysicalSET} and let $\phi$ be a solution to the wave equation associated with smooth compactly supported data. Consider also a timelike vector field $\tau$ that is transverse to the horizon and equal to $\partial_t$ for $r$ large enough~; the vector $\tau^a$ is of the form
\[ \tau^a \partial_a = \alpha \partial_t + \beta \frac{1}{F} (\partial_t - \partial_{r_*} ) \, ,\]
where $\alpha \geq 1$, $\alpha =1$ for $r$ large enough and $\beta \geq 0$, $\beta =0$ for $r$ large enough. Denote by $j_a$ the unrescaled energy current $1$-form associated with $\tau$,
\[ j_a = \tau^b T_{ab} \, .\]
The physical energy flux, associated with $\tau$, of the solution $\phi$ across $S_T$ is given by
\[ {\cal E}_{\tau ,S_T} = \int_{S_T} j_a N^a L \lrcorner \dvol_g \, ,\]
where $N^a$ is the normal vector field to $S_T$ associated via the metric $g$ to the co-normal $\d \Psi$,
\[ N_a \d x^a = \d \Psi \, ,~ N^a \frac{\partial}{\partial x^a} = g^{ab} N_b \frac{\partial}{\partial x^a} = F^{-1} \left( \frac{\partial}{\partial t} + \frac{r_*}{\sqrt{1+r_*^2}} \frac{\partial}{\partial r_*} \right) = \frac{1}{r^2} \hat{N}^a \frac{\partial}{\partial x^a}  \, , \]
and
\[ L^a \frac{\partial}{\partial x^a} = \hat{L}^a \frac{\partial}{\partial x^a} = \frac{1+r_*^2}{1+2r_*^2} \left( \frac{\partial}{\partial t} - \frac{r_*}{\sqrt{1+r_*^2}} \frac{\partial}{\partial r_*} \right) \, ,\]
so that $L_a N^a = g_{ab} L^a N^b = \hat{g}_{ab} \hat{L}^a \hat{N}^b = 1$. The energy flux ${\cal E}_{\tau ,S_T}$ decays as follows~:
\begin{equation} \label{31}
{\cal E}_{\tau ,S_T} \lesssim 1/T^2 \, .
\end{equation}
\item[(ii)] The solution also satisfies the following uniform decay estimates~:
\begin{equation} \label{32}
\sup_{S_T} \sqrt{r} \phi \lesssim 1/T\, ,~ \sup_{S_T} r \phi \lesssim 1/\sqrt{T} \, .
\end{equation}
\end{description}
\begin{remark}
The constants in front of the powers of $1/T$ in the estimates of Theorem 4.1 in \cite{DaRoLN} involve some higher order weighted energy norms (third order for (i) and sixth order for (ii)) of the data, which are all finite in our case. The details of these norms are not important to us here. We merely need to establish that for any smooth and compactly supported data, the energy of the rescaled field on $S_T$ tends to zero as $T\rightarrow +\infty$.
\end{remark}
\begin{proposition}
For smooth and compactly supported data $\phi$ and $\partial_t \phi$ at $t =0$ there exists $K>0$ such that for $T \geq 1$ large enough,
\[ {\cal E}_{\partial_t ,S_T} \leq \frac{K}{T} \, .\]
\end{proposition}
{\bf Proof.} First note that since $\alpha \geq 1$ and $\beta \geq 0$, thanks to the dominant energy condition, we have
\[ T_{ab} \tau^a N^b = \alpha T_{ab} (\partial_t)^a N^b + \beta T_{ab} (\partial_u)^a N^b \geq T_{ab} (\partial_t)^a N^b \, .\]
Hence the physical energy on $S_T$ associated with the vector field $\tau^a$ controls the physical energy on $S_T$ associated with the vector field $\partial_t$~:
\begin{equation} \label{EstEnDtTau}
{\cal E}_{\tau ,S_T} \geq {\cal E}_{\partial_t ,S_T} \, .
\end{equation}
Let us now compare the physical energy flux ${\cal E}_{\partial_t ,S_T}$ and the rescaled energy flux $\hat{\cal E}_{\partial_t ,S_T}$ using the relation $\hat\phi = r \phi$. First, we have
\begin{eqnarray*}
T_{ab} (\partial_t)^a N^b &=& \frac{1}{2F} \left( (\partial_t \phi )^2 + (\partial_{r_*} \phi )^2 + 2 \frac{r_*}{\sqrt{1+r_*^2}} \partial_t \phi \partial_{r_*} \phi + \frac{F}{r^2} \vert \nabla_{S^2} \phi \vert^2 \right) \\
&=& \frac{1}{2r^2F} \left( (\partial_t \hat\phi )^2 + F^2(\partial_{r} \hat\phi - \frac{\hat\phi}{r})^2 + 2 \frac{Fr_*}{\sqrt{1+r_*^2}} \partial_t \hat\phi (\partial_{r} \hat\phi - \frac{\hat\phi}{r}) + \frac{F}{r^2} \vert \nabla_{S^2} \hat\phi \vert^2 \right) 
\end{eqnarray*}
and since $L^a = \hat{L}^a$,
\[ L \lrcorner \dvol_g = r^4 \hat{L} \lrcorner \dvol_{\hat{g}} \, .\]
Therefore
\[ {\cal E}_{\partial_t , S_T} = \frac12 \int_{S_T} \left( (\partial_t \hat\phi )^2 + F^2(\partial_{r} \hat\phi - \frac{\hat\phi}{r})^2 + 2 \frac{Fr_*}{\sqrt{1+r_*^2}} \partial_t \hat\phi (\partial_{r} \hat\phi - \frac{\hat\phi}{r}) + \frac{F}{r^2} \vert \nabla_{S^2} \hat\phi \vert^2 \right) \frac{r^4}{r^2 F} \hat{L} \lrcorner \dvol_{\hat{g}} \]
and comparing with \eqref{FluxST}, we obtain
\begin{eqnarray*}
\hat{\cal E}_{\partial_t , S_T} &=& {\cal E}_{\partial_t , S_T} + \int_{S_T} \frac{M}{r} \hat\phi^2 \hat{L} \lrcorner \dvol_{\hat{g}} - \frac12 \int_{S_T} \left( F \frac{\hat\phi^2}{r^2} - 2  \frac{\hat\phi}{r} \partial_{r_*} \hat\phi - 2 \frac{r_*}{\sqrt{1+r_*^2}} \partial_t \hat\phi \frac{\hat\phi}{r} \right) r^2
 \hat{L} \lrcorner \dvol_{\hat{g}} \\
 &=& {\cal E}_{\partial_t , S_T} + \frac12 \int_{S_T} (\frac{2M}{r} -F) \hat\phi^2 \hat{L} \lrcorner \dvol_{\hat{g}} + \frac12 \int_{S_T}  2 \frac{\hat\phi}{r} \left( \partial_{r_*} \hat\phi +  \frac{r_*}{\sqrt{1+r_*^2}} \partial_t \hat\phi \right) r^2 \hat{L} \lrcorner \dvol_{\hat{g}} \\
& \leq & {\cal E}_{\partial_t , S_T} + \frac12 \int_{S_T} (\frac{2M}{r} -F) \hat\phi^2 \hat{L} \lrcorner \dvol_{\hat{g}} \\
&& + \frac12 \int_{S_T} \left( 2F \frac{\hat\phi^2}{r^2} +  \frac{1}{2F} \left( \partial_{r_*} \hat\phi +  \frac{r_*}{\sqrt{1+r_*^2}} \partial_t \hat\phi  \right)^2 \right)r^2 \hat{L} \lrcorner \dvol_{\hat{g}} \\
&\leq & {\cal E}_{\partial_t , S_T} + \frac12 \int_{S_T} (\frac{2M}{r} +F) \hat\phi^2 \hat{L} \lrcorner \dvol_{\hat{g}} \\
&& + \frac12 \int_{S_T} \frac{1}{2F} \left( ( \partial_{r_*} \hat\phi )^2+  \frac{2r_*}{\sqrt{1+r_*^2}} \partial_t \hat\phi \partial_{r_*} \hat\phi + ( \partial_{t} \hat\phi )^2 \right)r^2 \hat{L} \lrcorner \dvol_{\hat{g}} \\
&\leq & {\cal E}_{\partial_t , S_T} + \frac12 \int_{S_T} \hat\phi^2 \hat{L} \lrcorner \dvol_{\hat{g}} + \frac12 \hat{\cal E}_{\partial_t , S_T} \, .
\end{eqnarray*}
This gives us
\[ \hat{\cal E}_{\partial_t , S_T} \leq 2 {\cal E}_{\partial_t , S_T} + \int_{S_T} \hat\phi^2 \hat{L} \lrcorner \dvol_{\hat{g}} \, .\]
The second estimate in \eqref{32} says exactly that
\[ \sup_{S_T} \hat\phi^2 \lesssim \frac{1}{T} \, . \]
Since moreover
\[ \int_{S_T} \hat{L} \lrcorner \dvol_{\hat{g}} = \int_{\R \times S^2} \frac{F}{r^2} \sin \theta \d r_* \d \theta \d \varphi = \int_{[2M , +\infty [ \times S^2} \frac{1}{r^2} \sin \theta \d r \d \theta \d \varphi =\frac{2\pi}{M} <\infty \]
and ${\cal E}_{\partial_t , S_T} \lesssim 1/T^2$ by \eqref{31} and \eqref{EstEnDtTau}, this concludes the proof of the proposition. \qed

\begin{remark}
The finiteness of the last integral in the proof is strongly related to the finiteness of the volume of $S_T$ for the measure $\hat{\mu}_{S_T}$ induced by $\hat{g}$. As one can readily guess from the definitions of $S_T$ and $\hat{g}$, the volume of $S_T$ for the measure $\hat{\mu}_{S_T}$ is independent of $T$. Figure \ref{3surface} may be a little misleading in giving the impression that $S_T$ shrinks to a point, we must not forget that due to the way $\hat{g}$ is rescaled, $i^+$ is still at infinity. The volume of $S_T$ for $\hat{\mu}_{S_T}$ is easy to calculate. First we restrict $\hat{g}$ to $S_T$ using the explicit dependence of $t$ on $r_*$ on $S_T$~:
\[ \hat{g}\vert_{S_T} = - \left[ \frac{R^2 F}{1+r_*^2} \d r_*^2 +\d \omega^2 \right] \, .\]
Then we calculate $\hat{\mu}_{S_T}$~:
\[ \d \hat{\mu}_{S_T} = \frac{R\sqrt{F}}{\sqrt{1+r_*^2}} \d r_* \d^2 \omega = \frac{R}{\sqrt{F} \sqrt{1+r_*^2}} \d r \d^2 \omega\, .\]
So the volume of $S_T$ for $\hat{\mu}_{S_T}$ is
\[ \mathrm{Vol}_{\hat{g}} (S_T) = 4\pi \int_{2M}^{+\infty} \frac{1}{r\sqrt{F} \sqrt{1+r_*^2}} \d r = 4\pi \int_{2M}^{+\infty} \frac{\d r}{\sqrt{r^2-2Mr} \sqrt{1+r_*^2}} < +\infty \, .\]
Note that
\[ \d \hat{\mu}_{S_T} = \sqrt{\hat{g}_{ab} \hat{N}^a \hat{N}^b} \hat{L} \lrcorner \dvol_{\hat{g}} \, .\]
The two measures $\hat{\mu}_{S_T}$ and $\hat{L} \lrcorner \dvol_{\hat{g}}$ on $S_T$ are not uniformly equivalent since
\[ \sqrt{\hat{g}_{ab} \hat{N}^a \hat{N}^b} = \frac{r}{\sqrt{F} \sqrt{1+r_*^2}} \left\{ \begin{array}{ccl} { \simeq 1} & {\mbox{as}} & { r\rightarrow +\infty \, ,} \\ {\rightarrow +\infty} & {\mbox{as}} & { R \rightarrow 2M \, ,} \end{array} \right. \]
but this is integrable in the neighbourhood of $2M$.

If we had normalized $N$ to start with,
\[ \tilde{N}^a = \frac{1}{\sqrt{\hat{g}_{cd} \hat{N}^c \hat{N}^d}} \hat{N}^a \, ,\]
and put
\[ \tilde{L}^a = \sqrt{\hat{g}_{cd} \hat{N}^c \hat{N}^d} \hat{L}^a \, ,\]
so that $\hat{g}_{ab} \tilde{N}^a \tilde{L}^b =1$, then we would have
\[ \d \hat{\mu}_{S_T} = \tilde{L} \lrcorner \dvol_{\hat{g}} \, .\]
\end{remark}
So we have the following result~:
\begin{proposition} \label{PropEnergyIdentityFuture}
For smooth and compactly supported initial data on $\Sigma_0$, we have
\begin{equation} \label{EnergyIdentityFuture}
\hat{\cal E}_{\partial_t, \scri^+} + \hat{\cal E}_{\partial_t, \scrh^+} = \hat{\cal E}_{\partial_t, \Sigma_0} \, ,
\end{equation}
with
\begin{eqnarray*}
\hat{\cal E}_{\partial_t, \Sigma_0} &=& \int_{\Sigma_0} (\hat{J}_a+V_a)K^a \, r^2 F^{-1} \partial_t \lrcorner \dvol_{\hat{g}} \\
&=& \frac12 \int_{\Sigma_0} \left( (\partial_t \hat\phi )^2 + (\partial_{r_*} \hat\phi )^2 + R^2 F \vert \nabla_{S^2} \hat\phi \vert^2 + 2 MFR^3\hat\phi^2 \right) \d r_* \d^2 \omega \, ; \\
\hat{\cal E}_{\partial_t, \scri^+} &=& \int_{\scri^+} (\hat\nabla_K \hat\phi )^2 \hat{L}_{\scri^+} \lrcorner \dvol_{\hat{g}} = \int_{\scri^+} (\partial_u (\hat\phi \vert_{\scri^+} ) )^2 \d u \d^2 \omega\, ; \\
\hat{\cal E}_{\partial_t, \scrh^+} &=& \int_{\scrh^+} (\hat\nabla_K \hat\phi )^2 \hat{L}_{\scrh^+} \lrcorner \dvol_{\hat{g}} = \int_{\scrh^+} (\partial_v (\hat\phi \vert_{\scrh^+}) )^2 \d v \d^2 \omega \, .
\end{eqnarray*}
\end{proposition}
We can extend this result to minimum regularity initial data (i.e. data in $\cal H$) by standard density arguments, provided we give a meaning to the energy fluxes across $\scri$ and the horizon. We define a trace operator that to smooth and compactly supported initial data associates the future scattering data~:
\begin{definition}[Trace operator]
Let $(\hat\phi_0 , \hat\phi_1 ) \in {\cal C}^\infty_0 (\Sigma_0 ) \times {\cal C}^\infty_0 (\Sigma_0 )$. Consider the solution of \eqref{WEqResc} $\hat{\phi} \in {\cal C}^\infty ( \bar{\cal M} )$ such that
\[ \hat{\phi} \vert_{\Sigma_0} = \hat{\phi}_0 \, ,~ \partial_t \hat{\phi} \vert_{\Sigma_0} = \hat{\phi}_1 \, .\]
We define the trace operator ${\cal T}^+$ from ${\cal C}^\infty_0 (\Sigma_0 ) \times {\cal C}^\infty_0 (\Sigma_0 )$ to ${\cal C}^\infty (\scrh^+ ) \times {\cal C}^\infty (\scri^+ )$ as follows
\[ {\cal T}^+ (\hat\phi_0 , \hat\phi_1 ) = (\hat\phi \vert_{\scrh^+} , \hat\phi \vert_{\scri^+} ) \, .\]
\end{definition}
Then we extend this trace operator by density to $\cal H$ with values in the natural function space on $\scrh^+ \cup \scri^+$ inherited from \eqref{EnergyIdentityFuture}.
\begin{definition}[Function space for scattering data] \label{FuncSpaceScatt}
We define on $\scrh^+ \cup \scri^+$ the function space ${\cal H}^+$, completion of ${\cal C}^\infty_0 (\scrh^+ ) \times {\cal C}^\infty_0 (\scri^+ )$ in the norm
\begin{eqnarray*}
\Vert (\xi , \zeta ) \Vert_{{\cal H}^+} &=& \sqrt{ \int_{\scrh^+} \left( \hat\nabla_K \xi \right)^2 \hat{L}_{\scrh^+} \lrcorner \dvol_{\hat{g}} + \int_{\scri^+} \left( \hat\nabla_K \zeta \right)^2 \hat{L}_{\scri^+} \lrcorner \dvol_{\hat{g}}} \\
&=& \sqrt{ \int_{\scrh^+} \left( \partial_v \xi \right)^2 \d v \d^2 \omega + \int_{\scri^+} \left( \partial_u \zeta \right)^2 \d u \d^2 \omega} \, ,
\end{eqnarray*}
i.e.
\[ {\cal H}^+\simeq \dot{H}^1 (\R_v \, ;~ L^2 (S^2_\omega)) \times \dot{H}^1 (\R_u \, ;~ L^2 (S^2_\omega)) \, .\]
\end{definition}
\begin{remark}
The homogeneous Sobolev space (also referred to as the Beppo-Levi space) of order one on $\R$ is a delicate object. It is not a function space in the usual sense that its elements should belong to $L^1_\mathrm{loc}$, nor is it even a distribution space (see for example \cite{So1983} for a precise study of the one and two-dimensional cases). It is a space of classes of equivalence modulo constants. The reason is that constants have zero $\dot{H}^1$ norm and can in addition be approached in $\dot{H}^1$ norm by smooth and compactly supported functions on $\R$ (using a simple dilation of a given smooth compactly supported function whose value at the origin is the constant we wish to approach). The definition of $\dot{H}^1 (\R )$ by completion of ${\cal C}^\infty_0 (\R )$ in the $\dot{H}^1$ norm makes it the space of the limits of Cauchy sequences where indistiguishable limits are identified, i.e. a space of classes of equivalence modulo constants.

If one is reluctant to using classes of equivalence as scattering data, a more comfortable solution is to consider that the scattering data are in fact the traces of $\partial_t \hat{\phi}$ on $\scrh^+$ and $\scri^+$ and the function space in each case is then merely $L^2 (\R \times S^2 )$. This is what Friedlander did in is 1980 paper \cite{Fri1980}. It is however not clear to me that he did so for precisely this reason. He had, in my opinion, deeper motives for making this choice, guided as he was by the desire to recover the Lax-Phillips translation representer.

Whether one chooses to consider the scattering data as the traces of $\hat{\phi}$ (in $\dot{H}^1 (\R \, ;~ L^2 (S^2 ))$), or as the traces of $\partial_t \hat{\phi}$ (in $L^2 (\R \times S^2 )$) is purely a matter of taste, the two choices are equivalent.
\end{remark}
We infer from Proposition \ref{PropEnergyIdentityFuture} the following theorem~:
\begin{theorem} \label{ThmPartialIsometry}
The trace operator ${\cal T}^+$ extends uniquely as a bounded linear map from $\cal H$ to ${\cal H}^+$. It is a partial isometry, i.e. for any $(\hat{\phi}_0 , \hat{\phi}_1 ) \in {\cal H}$,
\[ \Vert {\cal T}^+ (\hat{\phi}_0 , \hat{\phi}_1 ) \Vert_{{\cal H}^+} = \Vert (\hat{\phi}_0 , \hat{\phi}_1 ) \Vert_{{\cal H}} \, .\]
\end{theorem}
An interesting property of second order equations is that once extended to act on minimal regularity solutions, the operator ${\cal T}^+$ can still be understood as a trace operator acting on the solution. We have seen in Proposition \ref{CauchyPb} that finite energy solutions of \eqref{WEqResc} belong to $H^1_\mathrm{loc} (\bar{\cal M})$ (see Remark \ref{Hsloc1} and Definition \ref{Hsloc2} for the definition of this function space). Elements of $H^1_\mathrm{loc} (\bar{\cal M})$ admit a trace at the conformal boundary that is locally $H^{1/2}$ on $\scrh^+ \cup \scri^+$. This is a consequence of a standard property of elements of $H^s (\Omega )$ for $\Omega$ a bounded open set of $\R^n$ with smooth boundary~; a function $f \in H^s (\Omega )$, $s > 1/2$, admits a trace on the boundary $\partial \Omega$ of $\Omega$ that is in $H^{s-1/2} (\partial \Omega )$. Hence the extended operator ${\cal T}^+$ is still a trace operator in a usual sense, i.e.
\[ {\cal T}^+ (\hat{\phi}_0 , \hat{\phi}_1 ) =  (\hat{\phi} \vert_{\scrh^+} , \hat{\phi} \vert_{ \scri^+ } ) \, .\]
This is in sharp contrast with what happens when working with first order equations like Dirac or Maxwell. In this case, finite energy solutions are in $L^2_\mathrm{loc} (\bar{\cal M})$ but in general not in $H^s_\mathrm{loc} (\bar{\cal M})$ for $s>0$. The density argument used in Theorem \ref{ThmPartialIsometry} would still give us an extension of the operator ${\cal T}^+$, whose range would be $L^2 (\scrh^+) \times L^2 (\scri^+ )$. This extended operator could not however be understood as a trace operator in the usual sense mentionned above, the regularity of the solutions being too weak.

\section{Scattering theory} \label{Scattering}

The construction of a conformal scattering theory on the Schwarzschild spacetime consists in solving a Goursat problem for the rescaled field on $\scrh^- \cup \scri^-$ and on $\scrh^+ \cup \scri^+$. In this section, we first solve the Goursat problem on $\scrh^+ \cup \scri^+$, the construction being similar in the past. Then we show that the conformal scattering theory entails a conventional analytic scattering theory defined in terms of wave operators. Since the exterior of a Schwarzschild black hole is static and the global timelike Killing vector $\partial_t$ extends as the null generator of $\scri^\pm$ and $\scrh^\pm$, it is easy to show that the past (resp. future) scattering data, i.e. the trace of the rescaled field on $\scrh^- \cup \scri^-$ (resp. on $\scrh^+ \cup \scri^+$) is a translation representer of the scalar field. We have a natural link between the conformal scattering theory and the Lax-Phillips approach, analogous to the one Friedlander established in his class of spacetimes. The difference is that in our case, the scattering data consist of a pair of data~: the trace of the rescaled field on null infinity (which is exactly the radiation field) and on the horizon.

\subsection{The Goursat problem and the scattering operator}

We solve the Goursat problem on $\scrh^+ \cup \scri^+$ following Hormander \cite{Ho1990}~: the principle is to show that the trace operator ${\cal T}^+$ is an isomorphism between $\cal H$ and ${\cal H}^+$. Theorem \ref{ThmPartialIsometry} entails that ${\cal T}^+$ is one-to-one and that its range is a closed subspace of ${\cal H}^+$. Therefore, all we need to do is to show that its range is dense in ${\cal H}^+$.

Let $(\xi , \zeta) \in {\cal C}^\infty_0 (\scrh^+) \times {\cal C}^\infty_0 (\scri^+)$, i.e. the support of $\xi$ remains away from both the crossing sphere and $i^+$ and the support of $\zeta$ remains away from both $i^+$ and $i^0$~; in other words, the values of $v$ remain bounded on the support of $\xi$ and the values of $u$ remain bounded on the support of $\zeta$. We wish to show the existence of $\hat\phi$ solution of \eqref{WEqResc} such that
\[ ( \hat{\phi } , \partial_t \hat{\phi} ) \in {\cal C} (\R_t \, ;~ {\cal H} ) \mbox{ and } {\cal T}^+ (\hat{\phi} \vert_{\Sigma_0} \, ,~ \partial_t \hat{\phi} \vert_{\Sigma_0} ) = (\xi , \zeta ) \, . \]
For such data the singularity at $i^+$ is not seen. We must however deal with the singularity at $i^0$. We proceed in two steps.

\begin{figure}[ht] %  figure placement: here, top, bottom, or page
\centering
\includegraphics[width=4in]{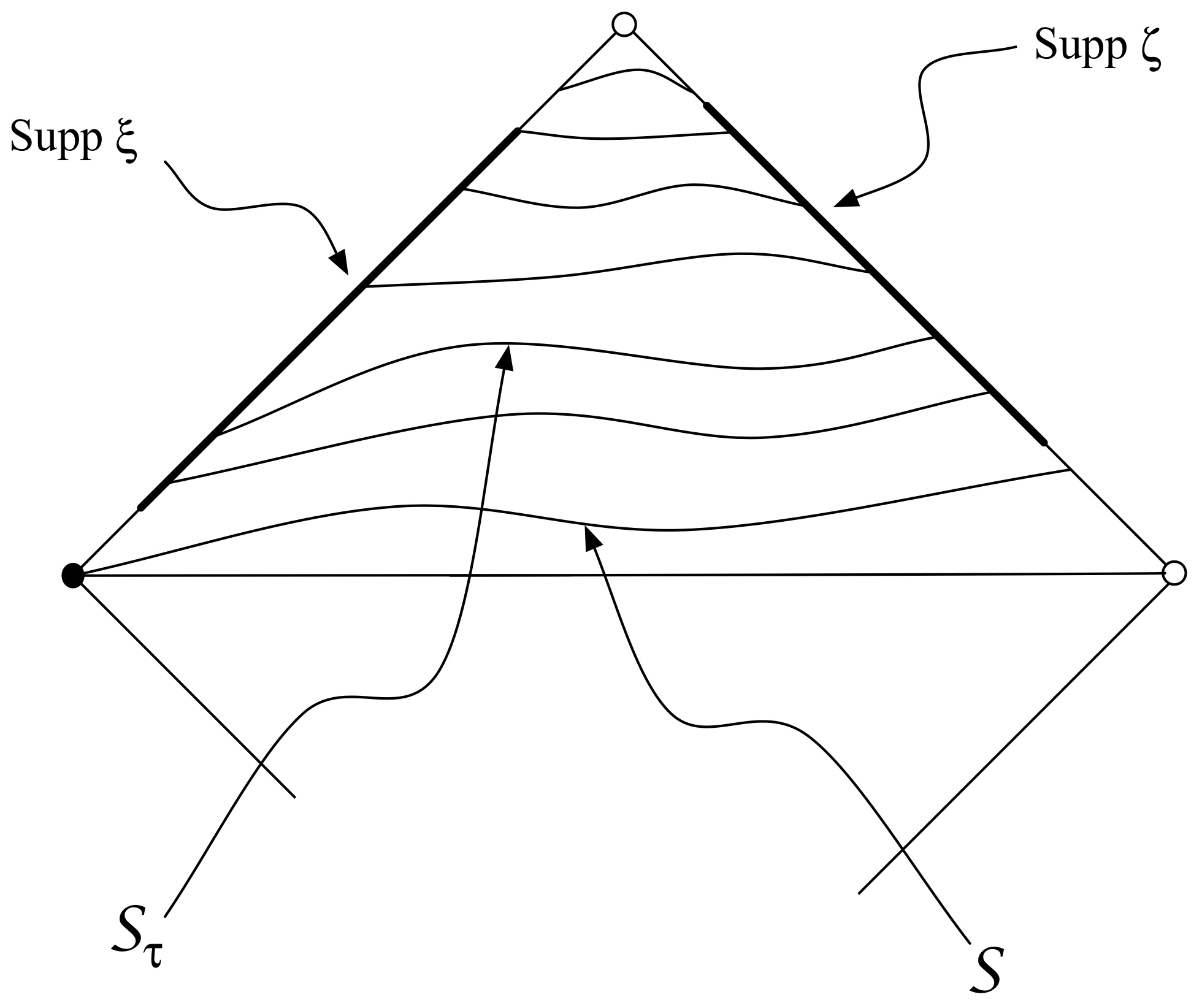}
\caption{A foliation $\{ {\cal S}_{\tau} \}_{\tau \geq 0}$ of $\overline{{\cal I}^+ ({\cal S})}$.} \label{Foliation}
\end{figure}

First, we consider $\cal S$ a spacelike hypersurface for $\hat{g}$ on $\bar{\cal M}$ that crosses $\scri^+$ in the past of the support of $\zeta$ and meets the horizon at the crossing sphere. The compact support of the data on $\scri^+$ allows us to apply the results of Hörmander \cite{Ho1990} even though we are not working with a spatially compact spacetime with a product structure (see Appendix \ref{HormGP} for details). We know from \cite{Ho1990} that there exists a unique solution $\hat{\Phi}$ of \eqref{WEqResc} such that~:
\begin{itemize}
\item $\hat{\Phi} \in H^1 ({\cal I}^+ ({\cal S}))$, where ${\cal I}^+ ({\cal S})$ is the causal future of $\cal S$ in $\bar{\cal M}$~; here we do not need to distinguish between $H^1 ({\cal I}^+ ({\cal S}))$ and $H^1_\mathrm{loc} ({\cal I}^+ ({\cal S}))$ because, due to the compact support of the Goursat data, the solution vanishes in a neighbourhood of $i^+$~;
\item given any foliation of ${\cal I}^+ ({\cal S})$ by $\hat{g}$-spacelike hypersurfaces $\{ {\cal S}_{\tau} \}_{\tau \geq 0}$, such that ${\cal S}_0 = {\cal S}$ (see figure \ref{Foliation}), $\hat{\Phi}$ is continuous in $\tau$ with values in $H^1$ of the slices and ${\cal C}^1$ in $\tau$ with values in $L^2$ of the slices~; in fact what we have a slightly stronger property, see Appendix \ref{HormGP} for a precise statement~;
\item $\hat{\Phi} \vert_{{\scri^+}} = \zeta$, $\hat{\Phi} \vert_{{\scrh^+}} = \xi$.
\end{itemize}

Second, we extend the solution down to $\Sigma_0$ in a manner that avoids the singularity at $i^0$. The crucial remark is that the restriction of $\hat{\Phi}$ to $\cal S$ is in $H^1 ({\cal S})$ and its trace on ${\cal S} \cap \scri^+$ is also the trace of $\zeta$ on ${\cal S} \cap \scri^+$, which is zero because of the way we have chosen $\cal S$. It follows that $\hat{\Phi} \vert_{\cal S}$ can be approached by a sequence $\{ \hat{\phi}^n_{0,{\cal S}} \}_{n\in \N}$ of smooth functions on $\cal S$ supported away from $\scri^+$ that converge towards $\hat{\Phi} \vert_{\cal S}$ in $H^1 ({\cal S})$. And of course $\partial_t \hat{\Phi} \vert_{{\cal S}}$ can be approached by a sequence $\{ \hat{\phi}^n_{1,{\cal S}} \}_{n\in \N}$ of smooth functions on $\cal S$ supported away from $\scri^+$ that converge towards $\partial_t \hat{\Phi} \vert_{{\cal S}}$ in $L^2 ({\cal S})$. Consider $\hat{\phi}^n$ the smooth solution of \eqref{WEqResc} on $\overline{\cal M}$ with data $( \hat{\phi}^n_{0,{\cal S}}\, ,~ \hat{\phi}^n_{1,{\cal S}})$ on $\cal S$. This solution vanishes in the neighbourhood of $i^0$ and we can therefore perform energy estimates for $\hat{\phi}^n$ between $\cal S$ and $\Sigma_0$~: we have the energy identity
\begin{equation} \label{EnIdentSSigma0}
{\cal E}_{\partial_t} ({\cal S} , \hat{\phi}^n ) = {\cal E}_{\partial_t} (\Sigma_0 , \hat{\phi}^n ) \, .
\end{equation}
\begin{remark}
The $H^1 \times L^2$ norm on $S$ is that induced by the rescaled metric $\hat{g}$. This is not equivalent to the norm induced by the energy ${\cal E}_{\partial_t}$ on $\cal S$, but the $H^1 \times L^2$ norm controls the other, which degenerates near null infinity and the crossing sphere. Consequently, $( \hat{\phi}^n_{0,{\cal S}}\, ,~ \hat{\phi}^n_{1,{\cal S}})$ is a Cauchy sequence also in the energy norm on $\cal S$.
\end{remark}
Similar energy identities between $\cal S$ and the hypersurfaces $\Sigma_t$ entail that $(\hat{\phi}^n , \partial_t \hat{\phi}^n )$ converges in ${\cal C} (\R_t \, ;~{\cal H} ) $ towards $(\hat{\phi} \, ,~ \partial_t \hat{\phi} )$, where $\hat{\phi}$ is a solution of \eqref{WEqResc}. By local uniqueness $\hat{\phi}$ coincides with $\hat{\Phi}$ in the future of $\cal S$. Hence if we denote
\[ \hat{\phi}_0 = \hat{\phi} \vert_{\Sigma_0} \, ,~ \hat{\phi}_1 = \partial_t \hat{\phi} \vert_{\Sigma_0} \, , \]
we have
\[ ( \hat{\phi}_0  \, ,~ \hat{\phi}_1 ) \in {\cal H} \]
and
\[ (\xi , \zeta ) = {\cal T}^+ ( \hat{\phi}_0  \, ,~ \hat{\phi}_1 ) \, .\]
This shows that the range of ${\cal T}^+$ contains ${\cal C}^\infty_0 (\scrh^+) \times {\cal C}^\infty_0 (\scri^+)$ and is therefore dense in ${\cal H}^+$. We have proved the following theorem.
\begin{theorem} \label{ThmGoursatPb}
The trace operator ${\cal T}^+$ is an isometry from $\cal H$ onto ${\cal H}^+$.
\end{theorem}
\begin{definition}
We introduce in a similar manner the past trace operator ${\cal T}^-$ and the space ${\cal H}^-$ of past scattering data\footnote{Note that the spaces ${\cal H}^\pm$ are naturally identified via a time reflexion $t \mapsto -t$.}. We define the scattering operator $S$ as the operator that to the past scattering data associates the future scattering data, i.e.
\[ S := {\cal T}^+ ({\cal T}^-)^{-1} \, .\]
The scattering operator is an isometry from ${\cal H}^-$ onto ${\cal H}^+$.
\end{definition}

\subsection{Wave operators} \label{WaveOps}

A conformal scattering construction such as the one we have just established can be re-interpreted as a scattering theory defined in terms of wave operators. This re-interpretation is more an a posteriori embellishment than a fundamental aspect of the theory, but it is interesting to realize that such fundamental objects of analytic scattering as wave operators, can be recovered from a purely geometrical construction which remains valid in time dependent geometries. To be completely precise, it is the inverse wave operators and the asymptotic completeness that we recover from the conformal scattering theory~; the direct wave operators are obtained in the classic analytic manner involving Cook's method. This choice is guided by simplicity and the flexibility of the method. The proof of existence of direct wave operators using Cook's method is the simplest part of analytic time-dependent scattering theory. Moreover, provided we have sufficiently explicit asymptotic information on our spacetime and good uniform energy estimates (without which we have in any case little hope of constructing a conformal scattering theory), it can be easily extended to fairly general non-stationary geometries, using a comparison dynamics that is defined geometrically, namely the flow of a family of null geodesics in the neighbourhood of the conformal boundary. The existence of inverse wave operators and asymptotic completeness, that we deduce from the conformal scattering construction in a direct manner, are the difficult aspects of analytic scattering.

When constructing wave operators using a conformal scattering theory, there is, just as for analytic scattering, some freedom in the choice of comparison dynamics, as well as some complications inherent to the fact that the full and simplified dynamics often act on different function spaces, defined on different manifolds that may not have the same topology.

The freedom of choice is two-fold. First we may choose different types of dynamics~: for the wave equation, we may wish to compare near infinity with the wave equation on flat spacetime or with a geometrically defined transport equation. In analytic scattering, the choice of comparison dynamics essentially fixes the space of scattering data as the finite energy space for the simplified Hamiltonian. In contrast, in conformal scattering, the energy space of scattering data is imposed by the energy estimates~; that is to say, the choice of vector field that we contract the stress-energy tensor with in order to get an energy current, fixes the functional framework, for both the scattering data and the initial data in fact. The comparison dynamics is then an additional choice, not completely determined by the space of scattering data. For instance, with a rather strong control on scattering data that seems to indicate the full flat spacetime wave equation as a natural simplified dynamics, we may yet choose a transport equation. All we really need is that the function space and the dynamics are compatible~: the comparison dynamics can usually be expressed as an evolution equation on the space of scattering data, whose coefficients are independent of the time parameter~; this compatibility then simply means that the Hamiltonian should be self-adjoint.

Second, for a given type of dynamics, there may still be some freedom. Say, if we choose a transport equation along a family of curves whose end-points span the conformal boundary, two different families of curves with the same end-points would work just as well.

In \cite{MaNi2004}, a conformal scattering construction on asymptotically simple spacetimes was re-interpreted as an analytic scattering theory defined in terms of wave operators. The comparison dynamics was determined by a null geodesic congruence in the neighbourhood of $\scri$, for which there are many choices. Also, some cut-off was required in a compact region in space, in order to avoid caustics. In the case we are considering here, the Schwarzschild geometry is sufficiently special that it singles out two congruences of null geodesics. Moreover, the topology of the spacetime (or equivalently the fact that the scattering data are specified on two disjoint null hypersurfaces instead of one in the asymptotically simple case) means that no cut-off is required.

The Schwarzschild spacetime is algebraically special of Petrov type D~; the four roots of the Weyl tensor are grouped at each point as two double principal null directions~: $\partial_t \pm \partial_{r_*}$. The two principal null congruences provide two preferred families of null curves along which to define a comparison dynamics.

We now proceed to introduce the full and the comparison dynamics as well as the other ingredients of the wave operators. We denote by ${\cal U} (t)$ the propagator for the wave equation on the finite energy space $\cal H$, i.e. for data $(\hat{\phi}_0 \, ,~ \hat{\phi}_1 ) \in {\cal H}$ at $t=0$, given $(\hat{\phi} \, ,~ \partial_t \hat{\phi} ) \in {\cal C} (\R_t \, ;~ {\cal H} )$ the associated solution of \eqref{WEqResc}, we have
\[ {\cal U} (t)  (\hat{\phi}_0 \, ,~ \hat{\phi}_1 )  = (\hat{\phi} (t) \, ,~ \partial_t \hat{\phi} (t)  ) \, .\]
The propagator ${\cal U} (t) $ is a strongly continuous one-parameter group of isometries on $\cal H$.
\begin{definition}
The comparison dynamics, denoted by ${\cal U}_0 (t)$, acts on pairs of functions on $\Sigma_0$ as the push-forward along the flow of the incoming principal null geodesics on the first function, and the push-forward along the flow of the outgoing principal null geodesics on the second function.

Considered as an operator on pairs of functions on the generic slice $\Sigma$, it acts as a translation to the left on the first function and a translation to the right on the second~:
\[ {\cal U}_0 (t) (\xi , \zeta ) (r_*,\omega) = \left( \xi (r_*+t , \omega ) , \zeta (r_*-t , \omega ) \right) \, .\]
It is a strongly continuous one-parameter group of isometries on
\begin{equation} \label{EnSpaceFree}
{\cal H}_0 = \dot{H}^1 (\R_{r_*} \, ;~ L^2 (S^2_\omega )) \times \dot{H}^1 (\R_{r_*} \, ;~ L^2 (S^2_\omega )) \, .
\end{equation}
\end{definition}
For our definition of direct and inverse wave operators, we need, in addition to the two dynamics ${\cal U} (t)$ and ${\cal U}_0 (t)$, an identifying operator, two cut-off functions and a pull-back operator between functions on the future conformal boundary and pairs of functions on $\Sigma_0$.
\begin{definition} \mbox{}
\begin{enumerate}
\item In order to obtain explicit formulae, we use on $\scrh^+$ the coordinates $(v,\omega)$, on $\scri^+$ the coordinates $(-u,\omega)$ and on $\Sigma_0$ we use $(r_* , \omega )$. Both for functions on $\scrh^+$ and $\scri^+$, we shall denote by $\partial_s$ the partial derivative with respect to their first variable, i.e. for $\xi$ a function on $\scrh^+$,
\[ \partial_s \xi = \partial_v \xi \]
and for a function $\zeta$ on $\scri^+$,
\[ \partial_s \zeta = - \partial_u \zeta \, .\]
\item We define the identifying operator
\[ {\cal J} \, :~ {\cal C}^\infty_0 (\Sigma ) \times {\cal C}^\infty_0 (\Sigma ) \rightarrow {\cal C}^\infty_0 (\Sigma ) \times {\cal C}^\infty_0 (\Sigma ) \]
by
\[ {\cal J} (\xi, \zeta ) (r_* , \omega ) = \left( \xi (r_* , \omega ) + \zeta (r_* , \omega ) \, ,~ \partial_s \xi (r_* , \omega ) - \partial_s\zeta (r_* , \omega ) \right) \, .\]
It combines pairs of functions on $\Sigma$ into initial data for equation \eqref{WEqResc}.
\item We also define two cut-off functions $\chi_\pm \in {\cal C}^\infty (\R) $, $\chi_+$ non decreasing on $\R$, $\chi_+ \equiv 0$ on $]-\infty , -1]$, $\chi_+ \equiv 1$ on $[1,+\infty [$, $\chi_- = 1-\chi_+$. They will be used with the variable $r*$ in order to isolate a part of the field living in a neighbourhood of either null infinity or the horizon. They can also be understood as functions on the exterior of the black hole~; we shall usually simply denote $\chi_\pm$ without referring explicitely to their argument.
\item We introduce the operator
\[ P^+ \,  : ~ {\cal C}^\infty_0 (\scrh^+) \times {\cal C}^\infty_0 (\scri^+) \longrightarrow  {\cal C}^\infty_0 (\Sigma_0) \times {\cal C}^\infty_0 (\Sigma_0) \]
that pulls back the first function along the flow of incoming principal null geodesics and the second along the flow of outgoing principal null geodesics. By the definition of the variables $u=t-r_*$ and $v=t+r_*$, in terms of coordinates $(r_*,\omega)$ on $\Sigma_0$, $(v,\omega )$ on $\scrh^+$ and $(-u,\omega)$ on $\scri^+$, the action of $P^+$ can be described very simply~: take $(\xi (v,\omega) , \zeta (-u,\omega)) \in {\cal C}^\infty_0 (\scrh^+) \times {\cal C}^\infty_0 (\scri^+)$,
\[ P^+ (\xi , \zeta ) (r_* , \omega ) = (\xi (r_*, \omega) , \zeta (r_*, \omega) ) \, .\]
The operator $P^+$ is an isometry from ${\cal H}^+$ onto ${\cal H}_0$ (see \eqref{EnSpaceFree} and Definition \ref{FuncSpaceScatt}).
\end{enumerate}
\end{definition}
\begin{remark}
The operator $P^+$ provides an identification between the conformal scattering data (that are functions defined on the conformal boundary) and initial data for the comparison dynamics (seen as acting between the slices $\Sigma_t$).
\end{remark}
\begin{theorem} \label{WaveOps}
The direct future wave operator, defined for smooth compactly supported scattering data
\[ (\xi , \zeta) \in {\cal C}^\infty_0 (\scrh^+ ) \times {\cal C}^\infty_0 (\scri^+ )\]
by
\[ W^+ (\xi , \zeta) := \lim_{t\rightarrow +\infty} {\cal U} (-t) {\cal J} \, {\cal U}_0 (t) P^+ (\xi , \zeta) \, ,\]
extends as an isometry from ${\cal H}^+$ onto $\cal H$.

The inverse future wave operator, defined for smooth compactly supported initial data for \eqref{WEqResc}
\[ (\hat{\phi}_0 , \hat{\phi}_1 ) \in {\cal C}^\infty_0 (\Sigma_0 ) \times {\cal C}^\infty_0 (\Sigma_0 ) \]
by
\[ \tilde{W}^+ (\hat{\phi}_0 , \hat{\phi}_1 ) = \lim_{t\rightarrow +\infty} (P^+)^{-1} \, {\cal U}_0 (-t) \left( \begin{array}{cc} \chi_- & 0 \\ \chi_+ & 0 \end{array} \right) {\cal U} (t) (\hat{\phi}_0 , \hat{\phi}_1 ) \, , \]
extends as an isometry from $\cal H$ onto ${\cal H}^+$.

Moreover, we have
\begin{gather}
\tilde{W}^+ = {\cal T}^+ = \slim_{t\rightarrow +\infty} (P^+)^{-1} \, {\cal U}_0 (-t) \left( \begin{array}{cc} \chi_- & 0 \\ \chi_+ & 0 \end{array} \right) {\cal U} (t) \, , \label{InvWOpSLim} \\
\tilde{W}^+ = (W^+)^{-1} \, .\label{InvWOp}
\end{gather}
\end{theorem}
\begin{remark}
It is important to understand that as soon as we have proved that $\tilde{W}^+ = {\cal T}^+$, we have established the asymptotic completeness, since ${\cal T}^+$ is an isometry from $\cal H$ onto ${\cal H}^+$. The proof of \eqref{InvWOpSLim} only relies on the conformal scattering construction. Once \eqref{InvWOpSLim} is established, all that remains to prove is the existence of the direct wave operator, which we do using Cook's method. The fact that $W^+$ is the inverse of $\tilde{W}^+$ is an immediate consequence of \eqref{InvWOpSLim} as we shall see. 
\end{remark}
\begin{remark}
The expressions of the wave operators can be simplified a little if we consider $\scri^+$ as the family of outgoing principal null geodesics and $\scrh^+$ as the family of incoming principal null geodesics. With this viewpoint, the comparison dynamics seen as acting on functions on $\scrh^+ \cup \scri^+$ reduces to the identity. We introduce a family of projections ${\cal P}_t$ that to a pair of functions $(\xi , \zeta) \in {\cal C}^\infty_0 (\scrh^+ ) \times {\cal C}^\infty_0 (\scri^+)$ associates its realization as a pair of functions on $\Sigma_t$, which as functions of $(r_*,\omega)$ have the following expression~:
\[ ( \xi (r_*+t,\omega ) , \zeta (r_*-t , \omega )) \, .\]
The direct and inverse wave operators acting on $(\xi , \zeta)$ then become~:
\begin{eqnarray*}
W^+ (\xi , \zeta) &=& \lim_{t\rightarrow +\infty} {\cal U} (-t) {\cal J}{\cal P}_t (\xi , \zeta) \, ;\\
\tilde{W}^+ (\hat{\phi}_0 , \hat{\phi}_1 ) &=& \lim_{t\rightarrow +\infty} {\cal P}_t^{-1} \left( \begin{array}{cc} \chi_- & 0 \\ \chi_+ & 0 \end{array} \right) {\cal U} (t) (\hat{\phi}_0 , \hat{\phi}_1 ) \, .
\end{eqnarray*}
We keep the version of the theorem however in order to get a closer similarity with the usual analytic expression of wave operators.
\end{remark}
{\bf Proof of Theorem \ref{WaveOps}.} All we need to do is prove that on a dense subspace of $\cal H$, $\tilde{W}^+$ is well-defined and coincides with ${\cal T}^+$, and that similarly, on a dense subspace of ${\cal H}^+$, $W^+$ is well-defined and coincides with $({\cal T}^+)^{-1}$.

Let us consider $(\hat{\phi}_0 , \hat{\phi}_1 ) \in {\cal C}^\infty_0 (\Sigma_0 ) \times {\cal C}^\infty_0 (\Sigma_0 ) \subset {\cal H}$. We denote by $\hat{\phi}$ the associated solution of \eqref{WEqResc} such that $(\hat{\phi} , \partial_t \hat{\phi}) \in {\cal C} (\R_t ; {\cal H})$ and put $(\xi , \zeta ) = {\cal T}^+ (\hat{\phi}_0 , \hat{\phi}_1 )$. For $t>0$, the operator
\[ (P^+)^{-1} \, {\cal U}_0 (-t) \left( \begin{array}{cc} \chi_- & 0 \\ \chi_+ & 0 \end{array} \right) {\cal U} (t) \]
first propagates the solution $\hat{\phi}$ up to the slice $\Sigma_t$, then cuts-off using $\chi_-$ (resp. $\chi_+$) the part of $\hat{\phi} (t)$ near infinity (resp. near the horizon) and puts the result in the first (resp. second) slot. Finally, the combination $(P^+)^{-1} \, {\cal U}_0 (-t)$ is the push-forward of the function in the first slot onto $\scrh^+$ along the flow of incoming principal null geodesics, and the push-forward of the function in the second slot onto $\scri^+$ along the flow of outgoing principal null geodesics.

\begin{figure}[ht] %  figure placement: here, top, bottom, or page
\centering
\includegraphics[width=4in]{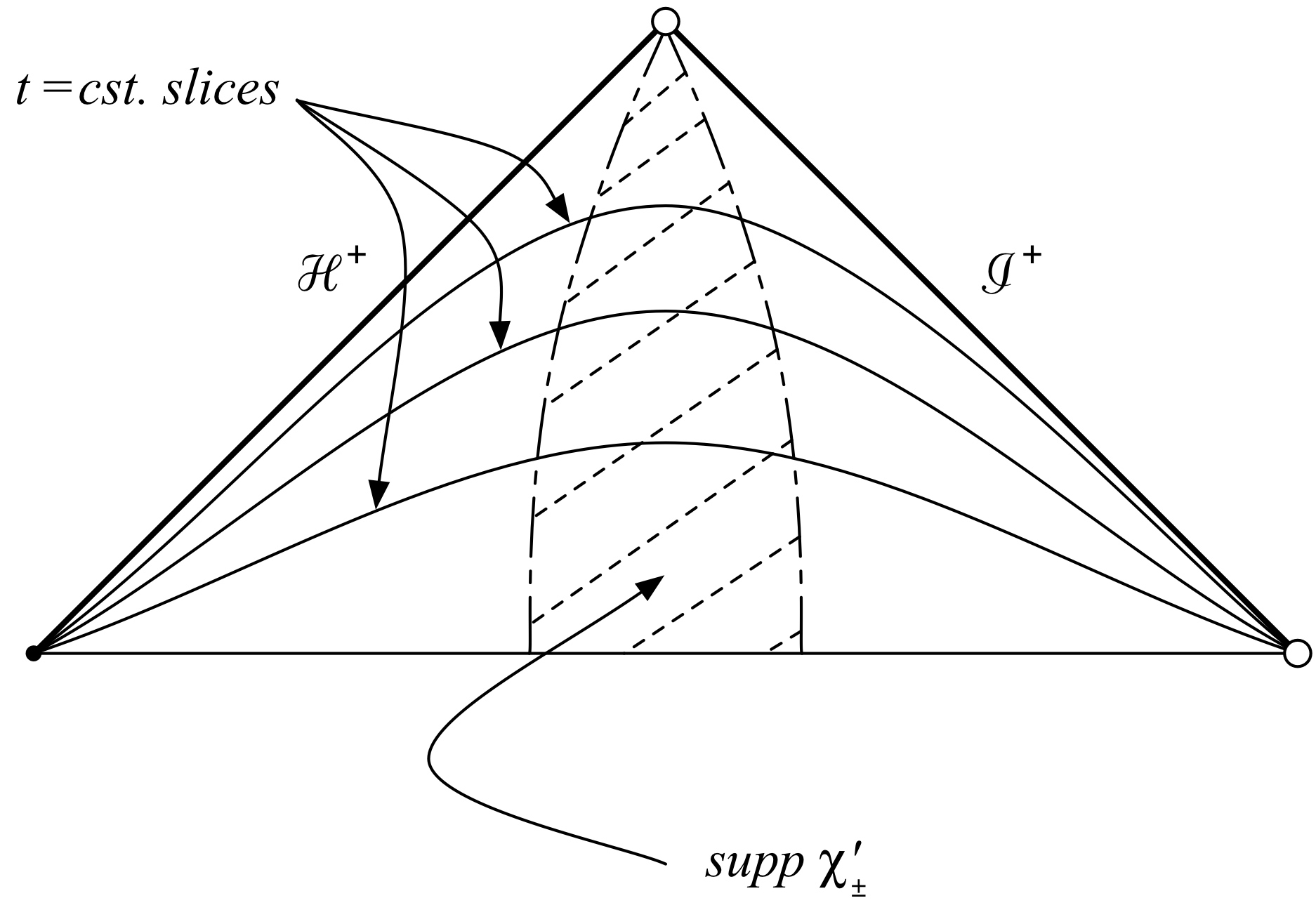}
\caption{The support of the derivatives of the cut-off functions $\chi_\pm$}  \label{SuppDerivativeCutOff}
\end{figure}

Since the support of the non constant part of the cut-off functions $\chi_\pm$ on $\Sigma_t$ remains away from both $\scri^+$ and $\scrh^+$ and accumulates at $i^+$ as $t \rightarrow +\infty$ (see figure \ref{SuppDerivativeCutOff}), we have the following pointwise limit
\[ \lim_{t\rightarrow +\infty} (P^+ )^{-1} {\cal U}_0 (-t) \left( \begin{array}{cc} \chi_- & 0 \\ \chi_+ & 0 \end{array} \right) {\cal U} (t) (\hat{\phi}_0 , \hat{\phi}_1 ) = (\xi , \zeta ) \, .\]
This already proves that $\tilde{W}^+$ is well-defined on smooth compactly supported initial data and coincides with ${\cal T}^+$ on this dense subset of $\cal H$. Therefore $\tilde{W}^+$ extends as the isometry ${\cal T}^+$ from $\cal H$ onto ${\cal H}^+$.

\begin{figure}[ht] %  figure placement: here, top, bottom, or page
\centering
\includegraphics[width=4in]{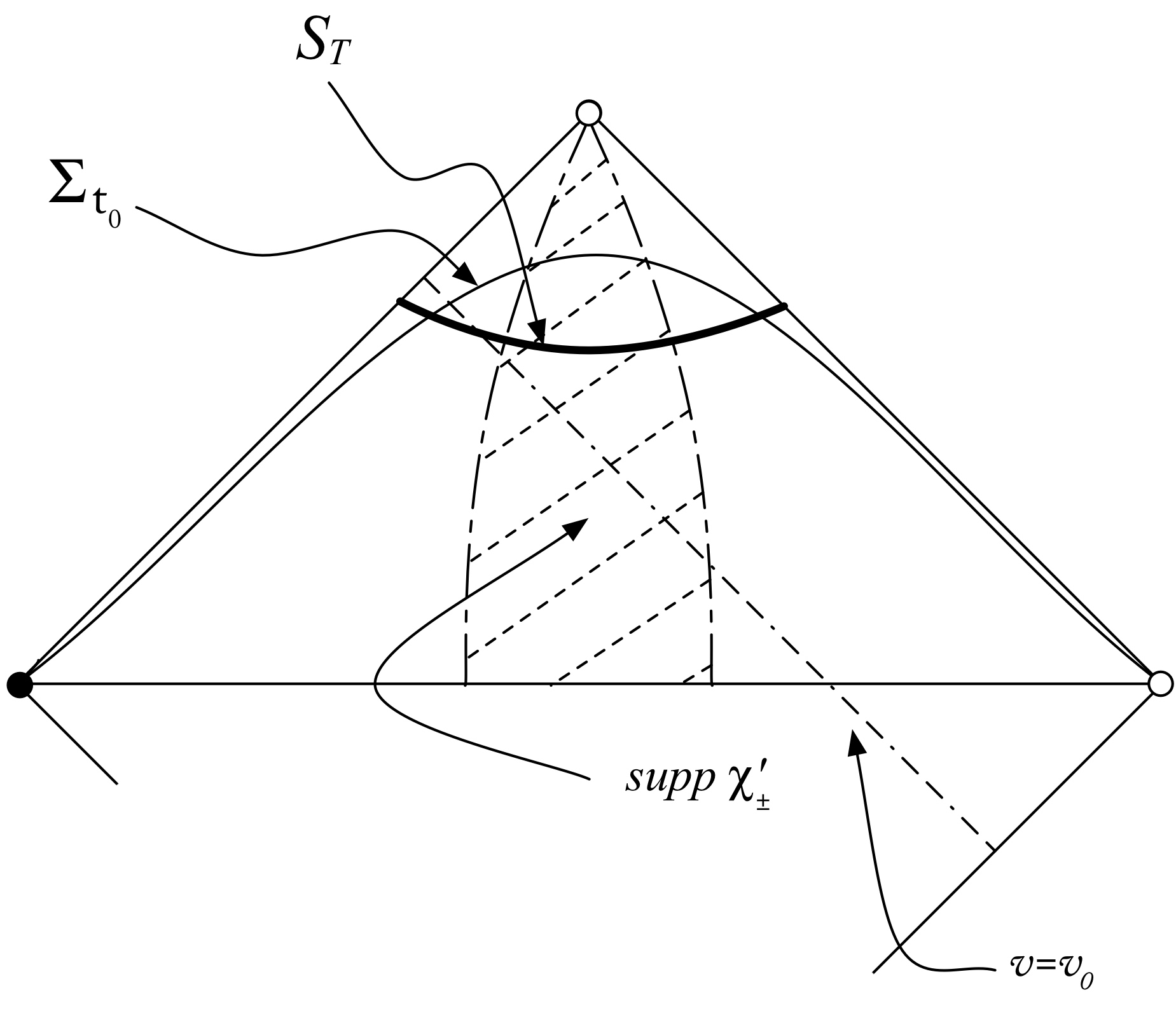}
\caption{The ingredients of the proof of \eqref{CvHorCInfty}.}
\label{StrongLimit}
\end{figure}

Let us now prove that the convergence above takes place in ${\cal H}^+$. This means that
\begin{eqnarray}
\lim_{t \rightarrow +\infty} \int_{\R \times S^2} \left( \frac{\partial}{\partial v} \left( \chi_- (-t+v)\hat{\phi} (t, -t+v, \omega) - \xi (v , \omega) \right) \right)^2 \d v \d \omega &=&0 \, , \label{CvHorCInfty} \\
\lim_{t \rightarrow +\infty} \int_{\R \times S^2} \left( \frac{\partial}{\partial u} \left( \chi_+ (t-u)\hat{\phi} (t, t-u, \omega) - \zeta (-u , \omega) \right) \right)^2 \d u \d \omega &=&0 \, . \label{CvInftyCInfty}
\end{eqnarray}
We prove \eqref{CvHorCInfty}, the proof of \eqref{CvInftyCInfty} is similar. Since $\hat{\phi} \in {\cal C}^\infty ( \bar{\cal M})$, we have
\[ \frac{\partial}{\partial v} \left( \chi_- (-t+v)\hat{\phi} (t, -t+v, \omega) - \xi (v , \omega) \right) \rightarrow 0 \mbox{ in } L^2_\mathrm{loc} (\R_{v} \, ;~ L^2 (S^2 )) \, .\]
In particular due to the compact support of the initial data, for any $v_0 \in \R$,
\begin{equation}
\lim_{t \rightarrow +\infty} \int_{]-\infty , v_0 [ \times S^2} \left( \frac{\partial}{\partial v} \left( \chi_- (-t+v)\hat{\phi} (t, -t+v, \omega) - \xi (v , \omega) \right) \right)^2 \d v \d \omega =0 \, .
\label{ConvCompact}
\end{equation}
Let $\varepsilon >0$, consider $T>0$ large enough such that $\hat{\cal E}_{\partial_t , S_T} < \varepsilon$. As a consequence, we also have that the energy flux across the part of $\scrh^+$ in the future of $S_T$ is lower than $\varepsilon$~:
\[ \hat{\cal E}_{\partial_t , (\scrh^+ \setminus \scrh^+_T)} < \varepsilon \, .\]
We choose $t_0 >0$ large enough such that for all $t>t_0$, the intersection of $\Sigma_t$ with the support of $\chi_-'$ is entirely in the future of $S_T$~; we also choose $v_0 >0$ such that the null hypersurface $\{ v=v_0 \}$ intersects all $\Sigma_t$, $t>t_0$, entirely in the future of $S_T$ (see figure \ref{StrongLimit} for an illustration of both choices). Then we have
\begin{eqnarray}
\int_{] v_0 , +\infty [ \times S^2} \left( \frac{\partial \xi}{\partial v} (v , \omega) \right)^2 \d v \d \omega &<& \varepsilon \, , \label{Small1}\\
\int_{] v_0 , +\infty [ \times S^2} \left( \frac{\partial}{\partial v} \left( \chi_- (-t+v)\hat{\phi} (t, -t+v, \omega) \right) \right)^2 \d v \d \omega &<& \varepsilon \, ,\hspace{0.1in} \mbox{ for all } t > t_0 \, . \label{Small2}
\end{eqnarray}
Now thanks to \eqref{ConvCompact}, we can choose $t_1 > t_0$ such that for all $t>t_1$ we have
\begin{equation} \label{Small3}
\int_{]-\infty , v_0 [ \times S^2} \left( \frac{\partial}{\partial v} \left( \chi_- (-t+v)\hat{\phi} (t, -t+v, \omega) - \xi (v , \omega) \right) \right)^2 \d v \d \omega < \varepsilon \, .
\end{equation}
Putting \eqref{Small1}, \eqref{Small2} and \eqref{Small3} together, we obtain that for $t>t_1$
\[ \int_{\R \times S^2} \left( \frac{\partial}{\partial v} \left( \chi_- (-t+v)\hat{\phi} (t, -t+v, \omega) - \xi (v , \omega) \right) \right)^2 \d v \d \omega < 5 \varepsilon \, . \]
This proves \eqref{CvHorCInfty} for data in ${\cal C}^\infty_0 (\Sigma_0 ) \times {\cal C}^\infty_0 (\Sigma_0 ) \subset {\cal H}$.

Let us now consider initial data $(\hat{\phi}_0 , \hat{\phi}_1 ) \in {\cal H}$. Still denoting $\hat\phi$ the associated solution of \eqref{WEqResc} and $(\xi , \zeta ) = {\cal T}^+ (\hat{\phi}_0 , \hat{\phi}_1 )$, we prove \eqref{CvHorCInfty} for such data. Let $\varepsilon >0$, consider $(\hat{\Phi}_0 , \hat{\Phi}_1 ) \in {\cal C}^\infty_0 (\Sigma_0 ) \times {\cal C}^\infty_0 (\Sigma_0 )$, $\hat{\Phi}$ the associated solution and $(\Xi , \mathrm{Z} ) = {\cal T}^+ (\hat{\Phi}_0 , \hat{\Phi}_1 )$, such that
\[ \Vert (\hat{\phi}_0 , \hat{\phi}_1 ) - (\hat{\Phi}_0 , \hat{\Phi}_1 ) \Vert_{{\cal H}}^2 < \varepsilon \, .\]
Then the energy fluxes, on $\scrh^+$ and $\Sigma_t$ for all $t$, of $\hat{\phi} - \hat{\Phi}$, are all lower than $\varepsilon$. Since \eqref{CvHorCInfty} is valid for $\hat\Phi$, we can find $t_0 >0$ such that for all $t >t_0$ we have
\[ \int_{\R \times S^2} \left( \frac{\partial}{\partial v} \left( \chi_- (-t+v)\hat{\Phi} (t, -t+v, \omega) - \Xi (v , \omega) \right) \right)^2 \d v \d \omega < \varepsilon \, . \]
It follows that for $t>t_0$, we have
\[ \int_{\R \times S^2} \left( \frac{\partial}{\partial v} \left( \chi_- (-t+v)\hat{\phi} (t, -t+v, \omega) - \xi (v , \omega) \right) \right)^2 \d v \d \omega < 9 \varepsilon \, . \]
This proves \eqref{CvHorCInfty} for finite energy data. We have therefore established \eqref{InvWOpSLim}.

Let us now consider $(\xi , \zeta ) \in {\cal C}^\infty_0 (\scrh^+ ) \times {\cal C}^\infty_0 (\scri^+ ) \subset {\cal H}^+$. For $t>0$, the operator
\[ {\cal U} (-t) {\cal J} \, {\cal U}_0 (t) P^+ \]
first (by the combination ${\cal U}_0 (t) P^+$) pulls back $\xi$ along the flow of incoming principal null geodesics and $\zeta$ along the flow of outgoing principal null geodesics, as a pair functions on $\Sigma_t$. Then ${\cal J}$ combines these two functions to obtain the initial data on $\Sigma_t$ for the wave equation~:
\[ \hat{\phi} \vert_{\Sigma_t} (r_*,\omega)= \xi (t+r_* , \omega) + \zeta (r_*-t , \omega) \, ,~  \partial_t \hat{\phi} \vert_{\Sigma_t} (r_* , \omega)= \partial_s \xi (t+r_* , \omega) - \partial_s \zeta (r_*-t , \omega)\, .\]
After which ${\cal U} (-t)$ propagates the corresponding solution of \eqref{WEqResc} down to $\Sigma_0$.

In order to prove that ${\cal U} (-t) {\cal J} \, {\cal U}_0 (t) P^+ (\xi , \zeta )$ converges in $\cal H$ as $t \rightarrow +\infty$, we use Cook's method~; the details of the proof can be found in Appendix \ref{AppendixCook}. Then it is easy to conclude that $W^+$ is the inverse of $\tilde{W}^+$. Let us consider for $(\xi , \zeta ) \in {\cal C}^\infty_0 (\scrh^+ ) \times {\cal C}^\infty_0 (\scri^+ )$ the quantity
\begin{equation} \label{CandidateInverse}
(P^+)^{-1} \, {\cal U}_0 (-t) \left( \begin{array}{cc} \chi_- & 0 \\ \chi_+ & 0 \end{array} \right) {\cal U} (t) {\cal U} (-t) {\cal J} \, {\cal U}_0 (t) P^+ (\xi , \zeta ) \, .
\end{equation}
By the strong convergence part of \eqref{InvWOpSLim} and the convergence in $\cal H$ of
\[ {\cal U} (-t) {\cal J} \, {\cal U}_0 (t) P^+ (\xi , \zeta ) \, , \]
\eqref{CandidateInverse} converges in ${\cal H}^+$ towards $\tilde{W}^+ W^+ (\xi , \zeta )$. But \eqref{CandidateInverse} simplifies as
\[ (P^+)^{-1} \, {\cal U}_0 (-t) \left( \begin{array}{cc} \chi_- & 0 \\ \chi_+ & 0 \end{array} \right) {\cal J} \, {\cal U}_0 (t) P^+ (\xi , \zeta ) = (P^+)^{-1} \, {\cal U}_0 (-t) \left( \begin{array}{cc} \chi_- & \chi_- \\ \chi_+ & \chi_+ \end{array} \right)  \, {\cal U}_0 (t) P^+ (\xi , \zeta ) \, .\]
Thanks to the compact support of $\xi$ and $\zeta$, this is equal to $(\xi , \zeta)$ for $t$ large enough. This concludes the proof. \qed
\begin{corollary}
Let us define similarly the past wave operators $W^-$ and $\tilde{W}^-$. We have
\[ \tilde{W}^- = (W^-)^{-1} ={\cal T}^- \, .\]
The scattering operator is related to the wave operators as follows
\[ S = \tilde{W}^+ W^- = (W^+)^{-1} W^- \, .\]
\end{corollary}

\subsection{Translation representer, scattering data, radiation field}

The conformal scattering theory we have constructed allows us, using the staticity of the exterior of a Schwarzschild black hole, to re-interpret immediately the scattering data as the crucial structure of the Lax-Phillips theory~: the translation representer. This is expressed in the following theorem.
\begin{theorem}
The scattering data are a translation representer of the associated scalar field. More precisely, consider $(\hat\phi , \partial_t \hat\phi ) \in {\cal C} (\R_t \, ;~ {\cal H} ) $ a solution to \eqref{WEqResc}, put $\hat\phi_0 := \hat\phi \vert_{\Sigma_0}$, $\hat\phi_1 := \partial_t \hat\phi \vert_{\Sigma_0}$ and
\[ (\xi , \zeta ) := {\cal T}^+ (\hat\phi_0 , \hat\phi_1 ) \, .\]
Then (expressing the functions using variables $(v,\omega)$ on $\scrh^+$ and $(-u,\omega )$ on $\scri^+$),
\[ {\cal T}^+ (\hat\phi \vert_{\Sigma_t}  , \partial_t \hat\phi \vert_{\Sigma_t} ) = (\xi (v+t , \omega ) , \zeta (-u-t , \omega )) \, .\]
\end{theorem}
{\bf Proof.} If instead of $(\hat\phi_0 , \hat\phi_1 ) $ we take $(\hat\phi \vert_{\Sigma_t}  , \partial_t \hat\phi \vert_{\Sigma_t} )$ for initial data, since $\partial_t$ is Killing, this is equivalent to pulling back the whole solution $\hat\phi$ of a time interval $t$ along the flow of $\partial_t$. Moreover $\partial_t$ extends as $\partial_v$ on $\scrh^+$ and as $\partial_u$ on $\scri^+$. This concludes the proof. \qed

Note that the part of the scattering data on $\scri^+$ is the trace of $\hat\phi = r \phi$ on $\scri^+$ and is therefore exactly the future radiation field. The essential difference from the theory of Lax-Phillips and the construction of Friedlander in 1980 \cite{Fri1980} is that we have a scattering theory with two scattering channels and therefore we need one extra scattering data. The important thing to understand here is that the translation representer is intimately related to the stationarity of the spacetime. If we give up stationarity, we also have to give up the translation representer but the conformal scattering construction would still be valid provided we have good estimates and a well-defined conformal boundary.

\section{Extension to the Kerr metric and concluding remarks} \label{Kerr}

At the time when this paper first appeared on the arXiv, the analysis in the Kerr framework was not as advanced as in the Schwarzschild setting. A variety of decay results were available for scalar waves and one for Maxwell fields, some of them establishing Price's law (which is the decay generically expected both in timelike directions and up the generators of null infinity, see R. Price \cite{Pri1972a} for scalar fields and \cite{Pri1972b} for zero-rest-mass fields)~: these results were due to L. Andersson and P. Blue \cite{ABlu}, M. Dafermos and I. Rodnianski \cite{DaRoLN}, F. Finster, N. Kamran, J. Smoller and S.T. Yau \cite{FiKaSmoYa1,FiKaSmoYa2}, F. Finster and J. Smoller \cite{FiSmo}, J. Metcalfe, D. Tataru and M. Tohaneanu \cite{MeTaTo}, D. Tataru and M. Tohaneanu \cite{TaTo} for the wave equation, and to L. Andersson and P. Blue \cite{ABlu2} for Maxwell fields. All these papers, except \cite{FiKaSmoYa1,FiKaSmoYa2,FiSmo}, deal with slowly rotating Kerr black holes. Decay estimates are useful in our conformal scattering construction in order to prove that the energy on the far future hypersurface $S_T$ tends to zero as $T \rightarrow +\infty$ (see subsection \ref{EnEstTInfinite}). This step however relies on the solidity of the foundations laid in subsection \ref{EnIdST}~: uniform energy estimates both ways, without loss, between a Cauchy hypersurface and the union of $S_T$ and the parts of $\scrh^+$ and $\scri^+$ in the past of $S_T$. Among the works we have just cited, the only one providing, if not exactly this kind of estimate, at least a way of obtaining them using the symmetry of the Kerr metric and the decay estimates, is \cite{ABlu}. They are the only ones establishing uniform estimates, for a positive definite energy, on a foliation by Cauchy hypersurfaces of the Kerr exterior. Many of the other works use the redshift effect near the horizon, see M. Dafermos and I. Rodnianski \cite{DaRo2009}. This is perfectly fine for proving decay, but the estimates cannot be reversed because when we go backwards in time, it is a blueshift effect that we have to deal with. The works of F. Finster, N. Kamran, J. Smoller and S.T. Yau rely on a different technique, which is an integral representation of the propagator for the wave equation~; they do not however obtain the type of estimate we need. The main drawback of the energy used by L. Andersson and P. Blue is that it is of too high order to be convenient for scattering theory. In fact, this is rather an aesthetic consideration than any serious scientific objection and it would be interesting to try an develop a conformal scattering theory based on their energy.

Since then, M. dafermos, I. Rodnianski and Y. Shlapentokh-Rothman \cite{DaRoShla} have obtained the missing uniform energy equivalence, without slow rotation assumption, and used it to construct a complete analytic scattering theory for the wave equation on the Kerr metric. They make the comment that it is crucial to chose an energy that does not see the redshift effect. Such an energy is based on a vector field that reduces at $\scrh^+$ to the null generator of the horizon, i.e. that is timelike outside the black hole but tangent to the horizon. This has interesting connections with our comments in section \ref{WaveOps}. It appears that with the results of \cite{DaRoShla}, our construction in the Schwarzschild case can now be extended to Kerr black holes essentially without change. It could be interesting to write this in detail provided we use only the relevent estimates and not the full scattering theory. Indeed, the re-interpretation of an analytic scattering theory as a conformal one is in many cases easy and purely a matter of understanding the scattering data as radiation fields (see A. Bachelot \cite{Ba1991} and D. Häfner and J.-P. Nicolas \cite{HaNi2004}). In the case of \cite{DaRoShla} the re-interpretation would be totally trivial since their scattering data are already described as radiation fields. The question of inferring an analytic scattering, defined in terms of wave operators, from a conformal scattering theory is more delicate however. It has been addressed in \cite{MaNi2004} and in the present work but still needs to be understood precisely in general. As far as the development of conformal scattering theory per se is concerned, it appears essential to find a way of replacing pointwise decay estimates, such as Price's law, by integrated decay estimates requiring a less precise knowledge of the local geometry and that are closer in nature to the minimal velocity estimates one obtains in the spectral approach to scattering theory (involving Mourre estimates and commutator methods).

\section{Acknowledgments}

I would like to thank Dean Baskin, Fang Wang and Jérémie Joudioux for interesting discussions that contributed to improve this paper. I am also indebted to the anonymous referee for his/her useful comments. This research was partly supported by the ANR funding ANR-12-BS01-012-01.

\appendix

\section{Cook's method for the direct wave operator} \label{AppendixCook}

In this proof we represent the free dynamics in a slightly different but equivalent manner. The space ${\mathbb H}_0 = \dot{H}^1 (\R_{r_*} \, ;~ L^2 (S^2 ) ) \times L^2 (\R_{r_*} \times S^2 )$ is the direct orthogonal sum of two supplementary subspaces~:
\[ {\mathbb H}_0^\pm = \{ (\psi_0 , \psi_1 ) \, ;~ \psi_1 = \mp \partial_{r_*} \psi_0 \} \, ;\]
via the operator $P^+$, ${\mathbb H}^+_0$ corresponds to $\dot{H}^1 (\R_{u} \, ;~ L^2 (S^2 ) )$ on $\scri^+$ and ${\mathbb H}^-_0$ to $\dot{H}^1 (\R_{v} \, ;~ L^2 (S^2 ) )$ on $\scrh^+$. On ${\mathbb H}_0$, we consider the free Hamiltonian
\[ H_0 = -i \left( \begin{array}{cc} 0 & 1 \\ {\partial_{r_*}^2 } & 0 \end{array} \right) \, .\]
The equation $\partial_t U = iH_0 U$ is the Hamiltonian form of the free equation
\[ \partial_t^2 \psi - \partial_{r_*}^2 \psi =0 \, .\]
The operator $H_0$ is self-adjoint on ${\mathbb H}_0$ and the free propagator ${\cal U}_0 (t)$ is just the group $e^{itH_0}$ conjugated by the identifying operator~:
\[ {\cal J} {\cal U}_0 (t) = e^{itH_0} {\cal J} \, .\]
With this description of the comparison dynamics, we need neither $P^+$ nor the identifying operator in the expression of the limit defining the direct wave operator $W^+$.

On $\cal H$ we consider the operator
\[ H = -i \left( \begin{array}{cc} 0 & 1 \\ {\partial_{r_*}^2 + \frac{F}{r^2} \Delta_{S^2} - \frac{FF'}{r}} & 0 \end{array} \right) \, ;\]
the equation $\partial_t U = iHU$ is the Hamiltonian form of \eqref{WEqResc}. The operator $H$ is self-adjoint on $\cal H$ and the propagator ${\cal U} (t)$ is equal to $e^{itH}$.

\begin{proposition}
For all $(U^h , U^\infty ) \in {\mathbb H}_0^- \times {\mathbb H}_0^+$, smooth and compactly supported, the following limits exist in $\cal H$~:
\begin{gather}
\lim_{t\rightarrow +\infty} e^{-itH} e^{itH_0} U^h \, , \label{LimHorizon} \\
\lim_{t\rightarrow +\infty} e^{-itH} e^{itH_0} U^\infty \, .\label{LimInfinity}
\end{gather}
\end{proposition}
{\bf Proof.} Take
\[ U^h = \left( \begin{array}{c} \psi_0 \\ \psi_1 = \partial_{r_*} \psi_0 \end{array} \right) \, ,~ \psi_0 \in {\cal C}^\infty_0 ( \R \, ;~ {\cal C}^\infty (S^2) ) \, .\]
A sufficient condition for the limit \eqref{LimHorizon} to exist is that
\[ \frac{\d }{\d t} e^{-itH} e^{itH_0} U^h = e^{-itH} \left( -i H + i H_0 \right) e^{itH_0} U^h \in L^1 (\R_t^+ \, ;~ {\cal H} ) \, .\]
Since $e^{-itH}$ is a group of unitary operators on $\cal H$, the condition is equivalent to
\[ \left( -i H + i H_0 \right) e^{itH_0} U^h \in L^1 (\R_t^+ \, ;~ {\cal H} ) \, .\]
This is easy to check~:
\[ \Vert \left( -i H + i H_0 \right) e^{itH_0} U^h \Vert^2_{{\cal H}} = \frac12 \int_{\R \times S^2} \left( -\frac{F}{r^2} \Delta_{S^2} \psi_0 (t+r_*) + \frac{FF'}{r} \psi_0 (t+r_*) \right)^2 \d r_* \d^2 \omega \]
and this falls off exponentially fast as $t \rightarrow +\infty$ thanks to the compact support and the smoothness of $\psi_0$ and to the fact that
\[ F(r) = 1 - \frac{2M}{r} = \frac{1}{r} e^{(r_*-r)/2M} \]
fall-off exponentially fast as $r_* \rightarrow -\infty$.

The proof of the existence of the other limit is similar, but we do not get exponential decay in this case. Take
\[ U^\infty = \left( \begin{array}{c} \psi_0 \\ \psi_1 = -\partial_{r_*} \psi_0 \end{array} \right) \, ,~ \psi_0 \in {\cal C}^\infty_0 ( \R \, ;~ {\cal C}^\infty (S^2) ) \, .\]
This time we have
\[ \Vert \left( -i H + i H_0 \right) e^{itH_0} U^\infty \Vert^2_{{\cal H}} = \frac12 \int_{\R \times S^2} \left( -\frac{F}{r^2} \Delta_{S^2} \psi_0 (r_*-t) + \frac{FF'}{r} \psi_0 (r_*-t) \right)^2 \d r_* \d^2 \omega \]
and this falls-off like $1/t^4$ as $t\rightarrow +\infty$, thanks to the compact support and the smoothness of $\psi_0$ and to the fact that
\[ \frac{F}{r^2} \simeq \frac{1}{r^2} \mbox{ and } r_* \simeq r \mbox{ at infinity.} \]
The other term falls off faster since
\[ \frac{FF'}{r} \simeq \frac{2M}{r^3} \mbox{ at infinity.} \]
So we still obtain the integrability in time of $\Vert \left( -i H + i H_0 \right) e^{itH_0} U^\infty \Vert_{{\cal H}}$ and this concludes the proof. \qed

As a consequence, for all $U_0 \in \mathbb{H}_0$, smooth and compactly supported, the limit
\[ \lim_{t\rightarrow +\infty} e^{-itH} e^{itH_0} U_0 \]
exists in $\cal H$. This is equivalent to the existence for smooth and compactly supported scattering data of the limit defining $W^+$.

\section{Applying L. Hörmander's results in the Schwarzschild setting} \label{HormGP}

\begin{figure}[ht!] %  figure placement: here, top, bottom, or page
\centering
\includegraphics[width=0.8\textwidth]{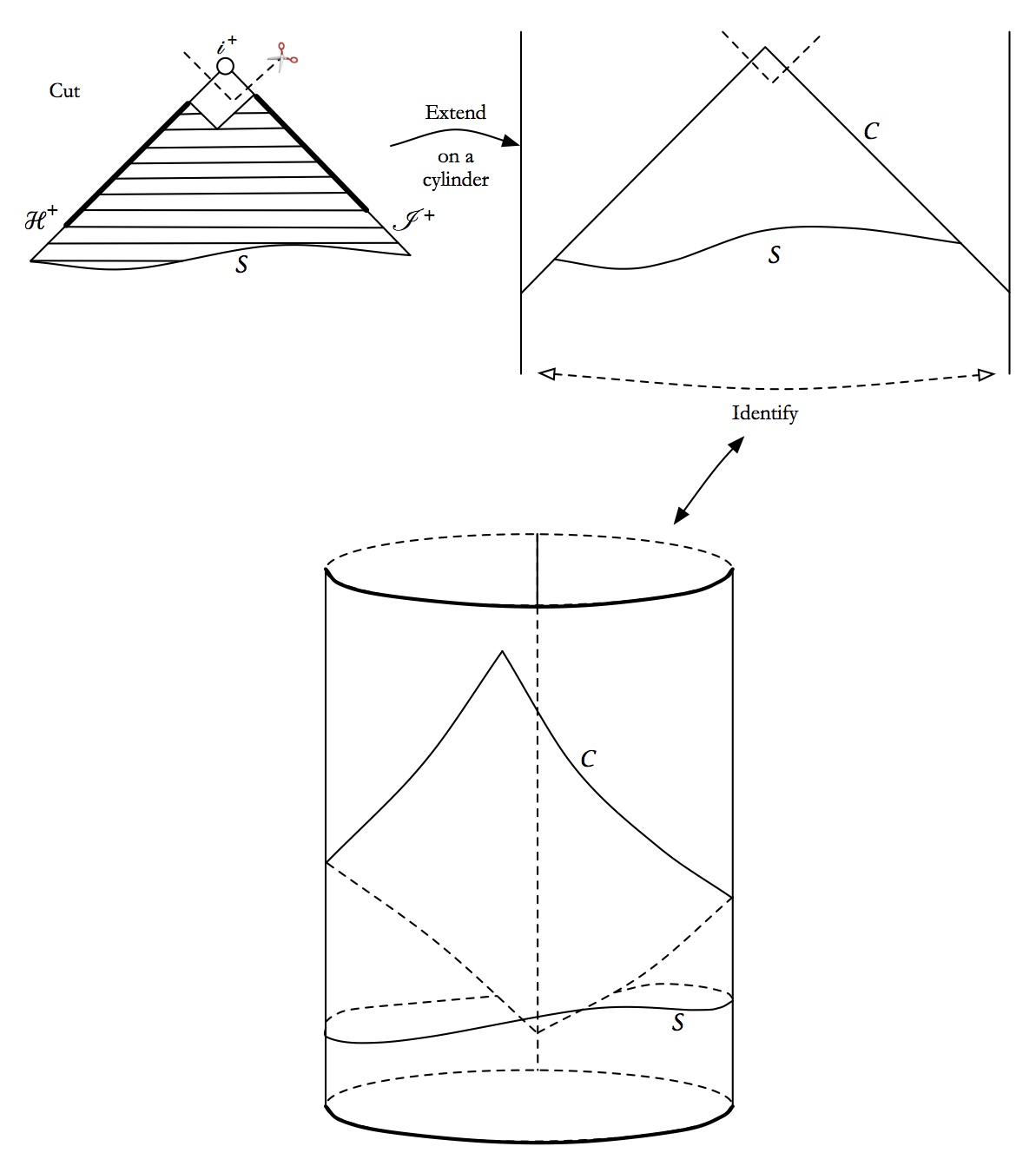}
\caption{A cut-extend construction for the solution of the Goursat problem from $\scri^+$.} \label{CutExtend}
\end{figure}
The work of L. Hörmander \cite{Ho1990} is a proof of the well-posedness of a weakly spacelike Cauchy problem, for a wave equation
\begin{equation} \label{ModWEq}
\partial_t^2 u - \Delta u + L_1 u = f \, ,
\end{equation}
on a Lorentzian spacetime that is a product $\R_t \times X$, with metric $\d t^2 - g$, $X$ being a compact manifold without boundary of dimension $n$ and $g(t)$ a Riemannian metric on $X$ smoothly varying with $t$. In \eqref{ModWEq}, $\Delta$ is a modified version of the Laplace-Beltrami operator in which the volume density associated with the metric is replaced by a given smooth density on $X$~; the operator $L_1$ is a first order differential operator with smooth coefficients and $f$ is a source. Any non degenerate change in the metric or the volume density can be absorbed in the operator $L_1$ so the results of \cite{Ho1990} are valid for the wave equation on any spatially compact globally hyperbolic spacetime. The data for the Cauchy problem are set on a hypersurface $\Sigma$ that is assumed Lipschitz and achronal (i.e. weakly spacelike), meaning that the normal vector field (which in the case of a Lipschitz hypersurface is defined almost everywhere) is causal wherever it is defined.

In the present work, we are not dealing with a spatially compact spacetime, but an easy construction allows us to understand the resolution of the Goursat problem for compactly supported data on the future conformal boundary as a Goursat problem on a cylindrical spacetime, for which Hörmander's results are valid.

The construction, described schematically in figure \ref{CutExtend}, goes as follows. The data on $\scrh^+ \cup \scri^+$ are compactly supported, which guarantees that the past of their support remains away from $i^+$. We simply consider the future ${\cal I}^+ ({\cal S})$ of the hypersurface $\cal S$ in $\bar{\cal M}$ (recall that $\cal S$ is a spacelike hypersurface on ${\bar{\cal M}}$ whose intersection with the horizon is the crossing sphere and which crosses $\scri^+$ strictly in the past of the support of the data) and we cut off the future $\cal V$ of a point in $\cal M$ lying in the future of the past of the support of the Goursat data (see figure \ref{CutExtend}). We denote by $\mathfrak{M}$ the resulting spacetime. Then we extend $\mathfrak{M}$ as a cylindrical globally hyperbolic spacetime $(\R_t \times S^3 \, ,~ \mathfrak{g})$. We also extend the part of $\scri^+ \cup \scrh^+$ inside ${\cal I}^+ ({\cal S}) \setminus {\cal V}$ as a null hypersurface $\cal C$ (see figure \ref{CutExtend}) that is the graph of a Lipschitz function over $S^3$ and the data by zero on the rest of the extended hypersurface. The Goursat problem for equation
\[ \square_\mathfrak{g} \psi + \frac16 \mathrm{Scal}_\mathfrak{g} \psi =0 \, ,\]
with the data we have just constructed has a unique solution (see \cite{Ho1990})
\[ \psi \in {\cal C} (\R \, ;~ H^1 (S^3 )) \cap {\cal C}^1 (\R \, ;~ L^2 (S^3 )) \, .\]
Then by local uniqueness and causality, using in particular the fact that as a consequence of the finite propagation speed, the solution $\psi$ vanishes in ${\cal I}^+ ({\cal S}) \setminus \mathfrak{M}$, the Goursat problem that we are studying has a unique smooth solution in the future of $\cal S$, that is the restriction of $\psi$ to $\mathfrak{M}$.

\end{document}